\documentclass[traditabstract,longauth]{aa}

\usepackage[varg]{txfonts}
\usepackage{epsfig}
\usepackage{graphicx}
\usepackage{float}
\usepackage{array}
\usepackage{longtable}

\begin{document}

\title{A Trio of GRB-SNe:}

\subtitle{GRB~120729A, GRB~130215A / SN~2013ez and GRB~130831A / SN~2013fu}

\author{Z. Cano\inst{1}
\and A. de Ugarte Postigo\inst{2}$^{,}$\inst{3}
\and A. Pozanenko\inst{4}
\and N. Butler\inst{5}
\and C. C. Th\"one\inst{2}
\and C. Guidorzi\inst{6}
\and T. Kr\"uhler\inst{3}$^{,}$\inst{7}
\and J.~Gorosabel\inst{2}$^{,}$\inst{8}$^{,}$\inst{9}
\and P. Jakobsson\inst{1}
\and G. Leloudas\inst{10}$^{,}$\inst{3}
\and D. Malesani\inst{3}
\and J. Hjorth\inst{3}
\and A. Melandri\inst{11}
\and C. Mundell\inst{12}
\and K.~Wiersema\inst{13}
\and P.~D'Avanzo\inst{11}
\and S. Schulze\inst{14}$^{,}$\inst{15}
\and A. Gomboc\inst{16}
\and A. Johansson\inst{1}
\and W. Zheng\inst{17}
\and D. A. Kann\inst{18}$^{,}$\inst{19}
\and F. Knust\inst{18}
\and K. Varela\inst{18}
\and C. W. Akerlof\inst{20}   
\and J. Bloom\inst{17}
\and O. Burkhonov\inst{21}
\and E. Cooke\inst{22}
\and J. A. de Diego\inst{23}
\and G. Dhungana\inst{24}
\and C.~Farina\inst{25}
\and F.~V.~Ferrante\inst{24}
\and H. A. Flewelling\inst{26}
\and O.~D.~Fox\inst{17}
\and J. Fynbo\inst{3}
\and N. Gehrels\inst{27}
\and L. Georgiev\inst{23}
\and J.~J.~Gonz\'alez\inst{23}
\and J.~Greiner\inst{18}
\and T.~G\"uver\inst{28}  
\and O. Hartoog\inst{29}
\and N. Hatch \inst{30}
\and M. Jelinek\inst{2}
\and R. Kehoe\inst{24}
\and S. Klose\inst{19}
\and E.~Klunko\inst{31}
\and D.~Kopa\v c\inst{16}
\and A. Kutyrev\inst{27}
\and Y.~Krugly\inst{32}
\and W. H. Lee\inst{23}
\and A. Levan\inst{33}
\and V. Linkov\inst{34}
\and A. Matkin\inst{35}
\and N. Minikulov\inst{36}
\and I. Molotov\inst{37}
\and J.~Xavier Prochaska\inst{38}
\and M.~G.~Richer\inst{39}
\and C. G. Rom\'an-Z\'u\~niga\inst{23}
\and V.~Rumyantsev\inst{40}
\and R. S\'anchez-Ram\'irez\inst{38}
\and I. Steele\inst{12}
\and N.~R.~Tanvir\inst{13}
\and A.~Volnova\inst{4}
\and A.~M.~Watson\inst{23}
\and D. Xu\inst{3}
\and F. Yuan\inst{41}
}

\offprints{ \email{zewcano@gmail.com}}

\institute{Centre for Astrophysics and Cosmology, Science Institute, University of Iceland, Reykjavik, Iceland. 
\and Instituto de Astrof\'isica de Andaluc\'ia (IAA-CSIC), Glorieta de la Astronom\'ia s/n, E-18008, Granada, Spain.  
\and Dark Cosmology Centre, Niels Bohr Institute, Juliane Maries Vej 30, Copenhagen \O, D-2100, Denmark.  
\and Space Research Institute, 117997, 84/32 Profsoyuznaya str., Moscow, Russia. 
\and School of Earth and Space Exploration, Arizona State University, Tempe, AZ 85287, USA. 
\and Department of Physics and Earth Sciences, University of Ferrara, via Saragat 1, I-44122 Ferrara, Italy. 
\and European Southern Observatory, Alonso de C\'ordova 3107, Vitacura, Casilla 19001, Santiago 19, Chile. 
\and Unidad Asociada Grupo Ciencia Planetarias UPV/EHU-IAA/CSIC, Departamento de F\'isica Aplicada I, E.T.S. Ingenier\'ia, Universidad del Pa\'is-Vasco UPV/EHU, Alameda de Urquijo s/n, E-48013 Bilbao, Spain.  
\and Ikerbasque, Basque Foundation for Science, Alameda de Urquijo 36-5, E-48008 Bilbao, Spain.  
\and The Oskar Klein Centre, AlbaNova, SE-106 91 Stockholm, Sweden.  
\and INAF/Brera Astronomical Observatory, via Bianchi 46, 23807, Merate (LC), Italy. 
\and Astrophysics Research Institute, Liverpool John Moores University, Liverpool Science Park, 146 Brownlow Hill, Liverpool, L3 5RF, UK. 
\and Department of Physics and Astronomy, University of Leicester, Leicester LE1 7RH, UK. 
\and Instituto de Astrof\'{i}sica, Facultad de F\'{i}sica, Pontificia Universidad Cat\'{o}lica de Chile, Av. Vicu\~{n}a Mackenna 4860, Santiago, Chile. 
\and Millennium Center for Supernova Science.  
\and Faculty of Mathematics \& Physics, University of Ljubljana,  Jadranska ulica 19, 1000 Ljubljana, Slovenia.  
\and Department of Astronomy, University of California, Berkeley, CA 94720-3411, USA.  
\and Max-Planck-Institut f\"ur extraterrestrische Physik, Giessenbachstra{\ss}e, 85748 Garching, Germany.  
\and Th\"uringer Landessternwarte Tautenburg, Sternwarte 5, 07778 Tautenburg, Germany.  
\and Department of Physics, University of Michigan, Ann Arbor, MI 48109, USA.   
\and Ulugh Beg Astronomical Institute, 100052, 33 Astronomicheskaya str., Tashkent, Uzbekistan. 
\and School of Physics and Astronomy, University of Nottingham, University Park, Nottingham, NG7 2RD, UK.  
\and Instituto de Astronom\'ia, Universidad Nacional Aut\'onoma de M\'exico, Apartado Postal 70-264, 04510 Mexico, D.F., Mexico.   
\and Department of Physics, Southern Methodist University, Dallas, Texas 75275, USA.  
\and Isaac Newton Group of Telescopes, Apartado de Correos 321, E-38700 Santa Cruz de La Palma, Canary Islands, Spain.  
\and Institute for Astronomy, University of Hawaii, 2680 Woodlawn Drive, Honolulu HI 96822, USA. 
\and NASA Goddard Space Flight Center, MS 665, Greenbelt, MD 20771, USA.  
\and Department of Astronomy and Space Sciences, Istanbul University Science Faculty, 34119 Istanbul, Turkey.  
\and Astronomical Institute, University of Amsterdam, Science Park 904, NL-1098 XH Amsterdam, the Netherlands.  
\and School of Physics and Astronomy, University of Nottingham, University Park, Nottingham, NG7 2RD, UK.  
\and Institute of Solar-Terrestrial Physics, Siberian department of the Russian Academy of Sciences, 664033, p/o box 291; Lermontov str., 126a, Irkutsk, Russia. 
\and Institute of Astronomy of Kharkiv National University, 61022, 35 Sumska str., Kharkiv, Ukraine.  
\and Department of Physics, University of Warwick, Coventry, CV4 7AL, UK.   
\and JSC Asronomical Scientific Center, Svobody str., 35, 22, 125362, Moscow, Russia.  
\and Ussuriysk Astrophysical Observatory, 692533, Primorskiy kray, Ussuriyskiy rayon, s. Gornotayozhnoe, Russia.   
\and Institute of Astrophysics, Academy of Sciences of Tajikistan, 734042, Bukhoro Str. 22, Dushanbe, Tajikistan.  
\and Keldysh Institute of Applied Mathematics, 125047, 4, Miusskaya sq., Moscow, Russia.  
\and Astronomy Department, University of California at Santa Cruz, CA 95064, USA.   
\and Instituto de Astronom\'ia, Universidad Nacional Aut\'onoma de M\'exico, Apartado Postal 106, 22800 Ensenada, Baja California, Mexico. 
\and Crimean Astrophysical Observatory, Taras Shevchenko National University of Kyiv, 98409, pgt. Nauchny, Crimea, Ukraine.  
\and Research School of Astronomy and Astrophysics, The Australian National University, Canberra, ACT 2611, Australia.  
}

\date{Received xx xxx 2014 / Accepted xx xxx 2014}

\abstract {We present optical and near-infrared (NIR) photometry for three gamma-ray burst supernovae (GRB-SNe): GRB~120729A, GRB~130215A / SN~2013ez and GRB~130831A / SN~2013fu.  In the case of GRB~130215A / SN~2013ez, we also present optical spectroscopy at $t-t_{0}=16.1$ d, which covers rest-frame 3000--6250 \AA.  Based on \ion{Fe}{ii} $\lambda5169$ and \ion{Si}{ii} $\lambda$6355, our spectrum indicates an unusually low expansion velocity of $\sim$4000--6350 km s$^{-1}$, the lowest ever measured for a GRB-SN.  Additionally, we determined the brightness and shape of each accompanying SN relative to a template supernova (SN~1998bw), which were used to estimate the amount of nickel produced via nucleosynthesis during each explosion.  We find that our derived nickel masses are typical of other GRB-SNe, and greater than those of SNe Ibc that are not associated with GRBs.  For GRB~130831A / SN~2013fu, we use our well-sampled \textit{R}-band light curve (LC) to estimate the amount of ejecta mass and the kinetic energy of the SN, finding that these too are similar to other GRB-SNe.  For GRB~130215A, we take advantage of contemporaneous optical/NIR observations to construct an optical/NIR bolometric LC of the afterglow.  We fit the bolometric LC with the millisecond magnetar model of Zhang \& M\'esz\'aros (2001), which considers dipole radiation as a source of energy injection to the forward shock powering the optical/NIR afterglow.  Using this model we derive an initial spin period of $P=12$ ms and a magnetic field of $B=1.1 \times 10^{15}$ G, which are commensurate with those found for proposed magnetar central engines of other long-duration GRBs.  }

\keywords{words. that are about keys. }

\maketitle

\section{Introduction}

Observational evidence supporting the connection between long-duration gamma-ray bursts (GRBs) and stripped-envelope, core-collapse supernovae (SNe) is now quite extensive (see  Woosley \& Bloom, 2006, and Hjorth \& Bloom 2012 for extensive reviews of gamma-ray burst supernovae; GRB-SNe).  2013 was a prosperous year for GRB-SN science, with no less than four spectroscopic GRB-SN associations: GRB~130215A / SN~2013ez (de. Ugarte Postigo et al. 2013b); GRB~130427A / SN~2013cq (de Ugarte Postigo et al. 2013c; Xu et al. 2013a; Levan et al. 2013; Melandri et al. 2014); GRB~130702A / SN~2013dx (Schulze et al. 2013) and GRB~130831A / SN2013fu (Klose et al. 2013; Nicuesa Guelbenzu et al. 2013).  These events join other spectroscopic GRB-SN associations (e.g. Galama et al. 1998; Hjorth et al. 2003; Stanek et al. 2003; Della Valle et al. 2003; Malesani et al. 2004; Pian et al. 2006; Chornock et al. 2010; Bufano et al. 2012; Berger et al. 2011; Sparre et al. 2011; Klose et al. 2012; de Ugarte Postigo et al. 2012; Melandri et al. 2012; Jin et al. 2013; Schulze et al. 2014).  Numerous photometric inferences of GRB-SNe via SN bumps in optical and near-infrared (NIR) light-curves (LCs; e.g. Zeh et al. 2004) further strengthen the GRB-SN connection (see Cano 2013 for a review).

The favoured physical description for producing a GRB is the ``collapsar'' scenario (Woosley 1993; MacFadyen \& Woosley 1999; MacFadyen et al. 2001), where a compact object forms during the collapse of a massive star and ejects shells of material collimated in a jet at relativistic velocities.  Multiple shells interact producing the initial $\gamma$-ray pulse, and as they propagate away from the explosion they encounter circumstellar material (CSM) ejected by the progenitor star prior to explosion (as well as interstellar material), producing a long-lived afterglow (AG).  In the simplest scenario, a forward shock (FS) is thought to be created when the shells interact with the CSM, which accelerate electrons that cool by emitting synchrotron radiation.  A couple of weeks (rest-frame) after the initial $\gamma$-ray pulse energetic SNe are then observed at optical and NIR wavelengths.

A basic assertion of the collapsar model is that the duration of the GRB prompt phase is the difference between the time that the central engine operates (i.e. $T_{90}$; though see Zhang et al. (2014) who argue that $T_{90}$ is not a reliable indicator of the engine activity timescale) minus the time it takes for the jet to breakout of the star: $T_{90} \sim t_{\rm engine}$ -- $t_{\rm breakout}$.  A direct consequence of this premise is that there should be a plateau in the distribution of $T_{90}$ for GRBs produced by collapsars when $T_{90}$ $<$ $t_{\rm breakout}$, which was confirmed by Bromberg et al. (2012).  Moreover, the value of $T_{90}$ found at the upper-limit of the plateau seen in three satellites (BATSE, \emph{Swift} and FERMI) was approximately the same ($T_{90}\sim$ 20--30 s), which is the typical breakout time of the jet.  This short breakout time suggests that the progenitor star at the time of explosion is very compact ($\sim 5$ $R_{\odot}$; Piran et al. 2013).  Bromberg et al. (2013) then used these distributions to calculate the probability that a given GRB arises from a collapsar or not based on its $T_{90}$ and hardness ratio.

The theoretical and observational evidence for the GRB-SNe connection is strong, however some questions remain unanswered.  One of the biggest uncertainties is the nature of the compact object that powers the GRB.  One possible engine is a stellar black hole rapidly accreting mass from a torus (e.g. Woosley 1993; MacFadyen et al. 2001).  Another scenario is the extraction of energy via the spin-down of a compact object through magnetic winds, either a neutron star with a very large magnetic field ($10^{15-16}$ Gauss) and rotating near breakup ($P\approx 1$ ms; i.e. a millisecond magnetar), or through the spin-down of a rapidly rotating (i.e. Kerr) black hole (e.g. van Putten et al. 2011).  Numerous flares and plateaus have been seen in AG LCs at X-ray and optical wavelengths (e.g. O'Brien et al. 2006, Margutti et al. 2010; Grupe et al. 2007, 2010), which require energy injection from a central engine.  The origin of the energy injection is still uncertain however, and may arise from different sources in different events.  

Secondly, the nature of the progenitor system has yet to be determined.  Due to the vast cosmological distances at which GRBs occur it is generally not possible to detect the progenitor directly, as has been done for the progenitors of other types of core-collapse SNe (e.g. Smartt et al. 2009; Maund et al. 2014).  Instead, the possible configuration of the progenitor system has to be indirectly inferred, whereby it is a formidable challenge to resolve the ambiguity between single and binary stars.  Arguments based on statistically significant sample sizes of the bolometric properties of GRB-SNe in relation to the other SN Ibc subtypes (Ib, Ic and Ic-BL; Cano 2013) indicate that the progenitors of most SNe Ibc likely arise from binary systems, where the mass of individual stars in the system is less than what is attributed to single Wolf-Rayet stars observed in nature (Crowther et al. 2007).  In these systems the outer layers of the star are tidally stripped, as well as ejected via line-driven winds.  

Conversely, the progenitors of SNe Ic-BL and GRB-SNe may arise from more massive single-star progenitors (e.g. Yoon \& Langer 2005; Woosley \& Heger 2006, Yoon et al. 2012), where rapid rotation mixes the interior of the star.  In these models a consequence of the rapid rotation is that the stars avoid a red supergiant stage, and there is a lack of strong coupling between the stellar core and surrounding envelope.  This allows more angular momentum to be retained in the core, and any mass loss experienced by the star (which also removes angular momentum) is reduced as it is lost from a much smaller surface area.  An expectation of these models is that the progenitors of SNe Ic-BL are more metal rich than those of GRB-SNe, (which has also been observed in nature, e.g. Modjaz et al. 2011; though see as well Levesque et al. 2012, Kr\"uhler et al. 2012, Savaglio et al. 2012 and Elliott et al. 2013 who have shown, respectively, that GRBs 020819A, 080605, 090323 and 110918A occurred in galaxies of solar and super-solar metallicities), and therefore lose more mass before exploding than GRB-SNe.  This provides a natural explanation for why a high-energy transient is observed in the latter because the central engine that is formed has retained more angular momentum at the time of explosion (e.g. Woosley \& Zhang 2007).  However, GRBs may also arise via binary systems, where the system may undergo a common-envelope phase.  If the system remains intact after one of the stars explodes, the in-spiral of the compact object into the core of the unexploded secondary can impart angular momentum to the core, which may be retained at the time of explosion to then power a GRB (e.g. Cantiello et al. 2007).

In this paper we attempt to address at least one of these outstanding questions, namely the nature of the compact object acting as a central engine of GRB~130215A.  Using the model of Zhang \& M\'esz\'aros (2001) we show that energy injection from a millisecond magnetar provides a plausible fit to an optical/NIR bolometric LC of the AG.  Using simple assumptions of the magnetar's mass and radius we derive physically plausible estimates of its magnetic field strength and initial spin period, finding they are consistent with those found for proposed magnetar central engines of other long-duration GRBs. 

The other main focus of this work is an investigation of the observational and physical properties of three GRB-SNe.  A key result we find is that SN~2013ez, associated with GRB~130215A, is more likely to be of type Ic than type IcBL, which all other GRB-SNe apart from SN~2002lt (associated with GRB~021211, Della Valle et al. 2003; see Section \ref{sec:Discussion}) have been spectroscopically classified as.  If our interpretation of the absorption features near $\sim5100$ and $\sim6200$ \AA $ $ as blueshifted \ion{Fe}{ii} $\lambda5169$ and \ion{Si}{ii} $\lambda6355$ (indicating blueshifted velocities of $\approx$4000 and $\approx6350$ km s$^{-1}$) is correct, it implies that SN~2013ez has the slowest ejecta velocities ever measured for a GRB-SN.  In sections \ref{sec:120729A}, \ref{sec:130215A} and \ref{sec:130831A} we present photometric observations of GRB~120729A, GRB~130215A / SN~2013ez and GRB~130831A / SN~2013fu\footnote{SN~2013fu was spectroscopically associated with GRB~130831A by Klose et al. 2013, in this work we present optical photometry only.}, respectively.  A SN signature is seen in each event, which arises via SN-bumps for GRBs 120729A and 130831A, and a bump+spectrum for GRB~130215A.  In section \ref{sec:Discussion} we discuss the observational and physical properties of these three GRB-SNe in relation to other SNe Ibc.  

Throughout this paper we use a $\Lambda$CDM cosmology constrained by Planck (Planck Collaboration et al. 2013) of $H_{0} = 67.3$ km s$^{-1}$ Mpc$^{-1}$, $\Omega_{M} = 0.315$, $\Omega_{\Lambda} = 0.685$.  Foreground extinction has been calculated using the dust extinction maps of Schlegel et al. (1998) and Schlafly \& Finkbeiner (2011), while values of the rest-frame extinction that have been derived from our data are presented in Table \ref{table:GRB_basic}.  All bolometric properties (nickel mass, ejecta mass and kinetic energy, $M_{\rm Ni}$, $M_{\rm ej}$ and $E_{\rm K}$, respectively) are calculated for the rest-frame filter range $UBVRIJH$ using the method in Cano (2013; C13 hereafter).  Unless stated otherwise, errors are statistical only.  Observer-frame times are used unless specified otherwise in the text.  The respective decay and energy spectral indices $\alpha$ and $\beta$ are defined by $f_{\nu} \propto (t - t_{0})^{-\alpha}\nu^{-\beta}$, where $t_{0}$ is the time at which the GRB triggered the BAT instrument on-board the \emph{Swift} satellite.


\begin{table*}
\centering
\setlength{\tabcolsep}{10.0pt}
\setlength{\extrarowheight}{3pt}
\caption{GRB-SNe: Vital Statistics}
\label{table:GRB_basic}
\begin{tabular}{cccccc}
\hline											
GRB	&	SN	&	$z$	&	$A_{V,\rm fore}$ (mag)	&	$A_{V,\rm rest}$ (mag)	&	$\rm d_{L}^{\dagger}$ (Mpc)	\\
\hline											
120729A	&	-	&	0.80	&	0.55	&	0.15	&	4910.7	\\
130215A	&	2013ez	&	0.597	&	0.53	&	0.0	&	3453.5	\\
130831A	&	2013fu	&	0.479	&	0.15	&	0.0	&	2664.2	\\
\hline	
\end{tabular}
\begin{flushleft}
$^{\dagger}$ Luminosity distance calculated using $H_{0} = 67.3$ km s$^{-1}$ Mpc$^{-1}$, $\Omega_{M} = 0.315$, $\Omega_{\Lambda} = 0.685$.
\end{flushleft}
\end{table*}

\section{GRB 120729A}
\label{sec:120729A}

GRB~120729A was detected at 10:56:14 UT on 29-July-2012 by the \emph{Swift} Burst Alert Telescope (BAT), and has a $T_{90}= 71.5\pm17.5$ s in the 15--350 keV energy range (Ukwatta et al. 2012; Palmer et al. 2012).  It was also detected by the \emph{Fermi} Gamma-Ray Burst Monitor (GBM) with a $T_{90}\approx 25$ s in the 50--300 keV energy range (Rau 2012).  Rapid follow-up by several ground-based telescopes identified an optical transient coincident with the XRT position (Virgili et al. 2012; Oates \& Ukwatta, 2012; Im \& Hong, 2012, Wren et al. 2012; Gorosabel et al. 2012; D'Avanzo et al. 2012), and a redshift of $z=0.80$ was measured with Gemini-North (Tanvir \& Ball, 2012).  The AG was not detected at radio (Laskar et al. 2012) down to $3\sigma$ upper limits of 39 mJy and 58 mJy, at 5.8 and 21.8 GHz, respectively.  The AG was also not detected at sub-mm wavelengths (350 GHz, with rms of 1.8 mJy; Smith et al. 2012)    An estimate of the isotropic energy release in $\gamma$-rays (1--10$^{4}$ keV, rest-frame) is $E_{\rm iso,\gamma}=2.3^{+0.3}_{-1.5} \times 10^{52}$ erg, while the peak energy $E_{\rm p} \approx 310.6$ keV\footnote{http://butler.lab.asu.edu/swift/bat$\_$spec$\_$table.html}.  The probability that GRB~120729A arises from a collapsar (Bromberg et al. 2013) based on $T_{90}$ alone is $99.996 \pm 0.001 \%$ (BAT) and $98.225 \pm 1.004 \%$ (GBM). We use a foreground extinction value of $E(B-V)_{\rm fore}$ = $0.164$ mag for GRB~120729A.

\subsection{Data Reduction \& Photometry}
\label{sec:120729A_datared}

We obtained observations with the 2-m Faulkes Telescope North (FTN) robotic telescope starting less than ten minutes after the $\gamma$-ray detection.  Subsequent follow-up observations were obtained with the 2-m Liverpool Telescope (LT), the 0.82-m Instituto de Astrof\'isica de Canarias (IAC) IAC80 telescope, the 3.6-m Telescopio Nazionale Galileo (TNG), and the 10.4-m Gran Telescopio Canarias (GTC) telescope. Six epochs of GTC images in $griz$ were obtained during the first month, and a final epoch in all filters at $t-t_{0}\approx 190$ d that were used as templates for image subtraction.  Image reduction of data obtained on all telescopes was performed using standard techniques in IRAF\footnote{IRAF is distributed by the National Optical Astronomy Observatory, which is operated by the Association of Universities for Research in Astronomy, Inc., under cooperative agreement with the National Science Foundation.}.

Calibration of the GTC data has been performed using standard star photometry.  Observations (in $griz$) of Landolt standard field PG1323-086 (Landolt 1992) were obtained the same night as the final GTC epoch, all of which were taken under photometric conditions.  The $BVR_{c}I_{c}$ magnitudes of PG1323-086 were transformed into $griz$ using transformation equations from Jordi et al. (2006), and the subsequent calibration was done using a zeropoint between the instrumental and catalog magnitudes.  The calibration in each filter was then used to create a set of secondary standards in the GRB field, which the GTC images are calibrated against.  

The $gVR_{c}i$ FTN, LT and TNG images are shallower than the GTC ones, where common stars are either saturated in the GTC images or not visible in the FTN/LT ones.  Instead these images have been independently calibrated via standard star photometry using Landolt standards taken with the TNG the same night as the GRB observations.  A zeropoint was computed between the Landolt standards and the instrumental $gVR_{c}i$ magnitudes, which was then used to create a small set of secondary standards in the GRB field that were visible in all FTN/LT/TNG images but not saturated.  The $R_{c}$ magnitudes were then transformed into $r$ magnitudes using transformation equations from Lupton (2005)\footnote{http://www.sdss.org/dr4/algorithms/sdssUBVRITransform.html}, which requires colours between filters (e.g. $R_{c}-i$) in order to properly calculate the corresponding SDSS magnitudes.  Early observations in $R_{c}$ and $i$ were taken within 5--10 minutes of each other, however, as these observations were taken very soon after the initial GRB trigger (after only a few tens of minutes), it may not be appropriate to assume no colour/spectral evolution during the timing of a given $R_{c}$-band and $i$-band observation.  Instead we estimated magnitudes in $r$ at the time of the $R_{c}$-band observation by two methods: (1) a log-linear interpolation of the $i$-band LC, and (2) fitting a broken PL to the $i$-band LC in order to extrapolate to earlier and later times.  Once the fitted functions were determined, the $i$ magnitude at the time of the $R_{c}$ observation was outputted.  When possible we have taken the average magnitude found with both of these methods, and included their standard deviation when calculating the SDSS magnitude in $r$.

We used our deep GTC images to obtain image-subtracted magnitudes of the optical transient (OT) associated with GRB~120729A, using the final epoch in each filter as a template.  This method proved to be valuable to isolate the OT flux as the field is quite crowded, and because the OT is very faint.  Already at $t-t_{0}=0.75$ d the $griz$ magnitudes of the OT are $m_{AB}=$24--25.  Image subtraction was performed using an adaptation of the original ISIS program (Alard \& Lupton 1998; Alard 2000) that was developed for Hubble Space Telescope SN surveys by Strolger et al. (2004).  A key advantage of this code is the option for the user to specify a set of stamps for the program to use when it calculates the point-spread function in each image.  The image-subtraction technique was then optimised by varying the kernel mesh size and measuring the standard deviation ($\sigma$) of the background counts in a nearby region in the image (where images with lower $\sigma$ values indicate they are a better subtracted image).  As a self-consistency check, we compared the OT magnitudes against those found by performing photometry on the un-subtracted images, converting the magnitudes into fluxes, and then mathematically subtracting the host flux away.  Good agreement was obtained with both methods, showing that the image-subtraction technique was well optimised.

The $griz$ magnitudes of the host galaxy were measured, and these magnitudes were converted into monochromatic fluxes using the flux zeropoints from Fukugita et al. (1995) and subtracted from the earlier observations obtained with the LT, FTN and IAC80.  The apparent magnitudes (not corrected for foreground or host extinction) of the GRB+SN+host are presented in Table \ref{table:photometry}.  

\subsection{The Afterglow}

\begin{figure}
 \centering
 \includegraphics[bb=0 0 263 189, scale=0.93]{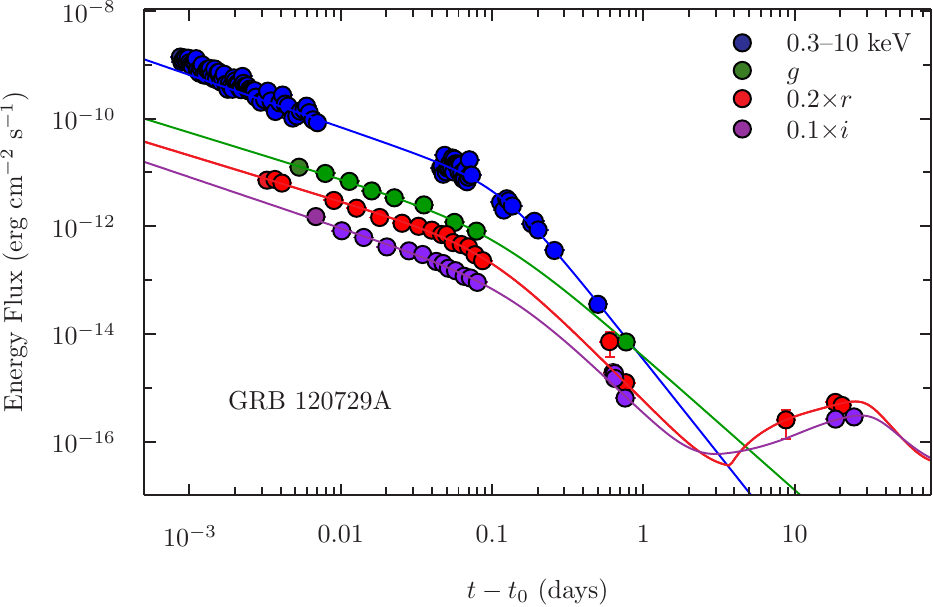}
 \caption{GRB 120729A: Optical and X-ray (0.3--10 keV) light curves.  $BR_{c}I_{c}$ magnitudes have been transformed into $gri$ using transformation equations from Jordi et al. (2006), see the main text for details.  The optical data are host-subtracted and have been corrected for foreground and rest-frame extinction.  Note that the errorbars are usually smaller than the size of the plotted symbols.  All LCs have been fit with a broken power-law in order to determine the decay rate before ($\alpha_{\nu,1}$) and after ($\alpha_{\nu,2}$) and the break ($T_{\nu,\rm B}$), as well as the timing of the break.  It is seen that $\alpha_{1}$ is approximately the same in the optical and X-ray, as well as the time of the break ($T_{\rm B}\approx 0.1$ d).  After the break the X-ray decays at a faster rate than the optical filters.  In $r$ and $i$ we simultaneously fit a SN-component (i.e. a stretch and luminosity factor relative to a redshifted, k-corrected template LC).  The paucity of optical points limits our analysis, however when fixing the stretch factor to $s=1.0$ in both filters, we find luminosity factors of $k_{r}=1.29\pm0.19$ and $k_{i}=0.76\pm0.11$.  }
 \label{fig:120729A_LC}
\end{figure}

We combined our optical detections with the \emph{Swift} XRT (0.3--10 keV) observations (Fig. \ref{fig:120729A_LC}), where the host-subtracted magnitudes are corrected for foreground and rest-frame extinction, and been converted into monochromatic fluxes (mJy), and then into energy fluxes (erg cm$^{-2}$ s$^{-1}$) using the zeropoints and filter effective wavelengths from Fukugita et al. (1995).

We fit all LCs with a broken power-law (PL) (we also included a SN-component that is simultaneously determined when fitting the $r$ and $i$ LCs, see section \ref{sec:120729A_SN}) in order to determine the decay rate before ($\alpha_{\nu,1}$) and after ($\alpha_{\nu,2}$) the break, as well as the timing of the break ($T_{\nu,\rm B}$).  Our best-fitting parameters (fit between 0.005--30 d) are: (1) X-ray: $\alpha_{X,1}=0.97\pm0.06$, $\alpha_{X,2}=3.54\pm0.27$, $T_{X,\rm B}=0.12\pm0.02$ d; (2) optical: $\alpha_{g,1}=0.86\pm0.04$, $\alpha_{g,2}=2.49\pm0.14$, $T_{g,\rm B}=0.11\pm0.02$ d; $\alpha_{r,1}=0.86\pm0.03$, $\alpha_{r,2}=2.85\pm0.10$, $T_{r,\rm B}=0.10\pm0.02$ d; $\alpha_{i,1}=0.95\pm0.07$, $\alpha_{i,2}=2.77\pm0.22$, $T_{i,\rm B}=0.14\pm0.04$ d.  The time the LC breaks is approximately the same at all frequencies ($T_{\rm B}\approx 0.11$ d).  The value of $\alpha_{1}$ is roughly the same at all wavelengths before the break, and while $\alpha_{2}$ is steeper in the X-ray than the optical, it is quite similar in all optical bands, though there is a hint that it is decay slightly slower in $g$, though of course the paucity of observations limits how much we can comment on this.  If instead we fit the optical LCs simultaneously, assuming that the time the LC breaks and the decay constants before and after the break are the same, we find $\alpha_{\rm opt,1}=0.89\pm0.09$, $\alpha_{\rm opt,2}=2.70\pm0.18$, $T_{\rm opt, B}=0.10\pm0.04$ d.


If the achromatic break at $t-t_{0}\approx0.11$ d is interpreted as a jet break, it is possible to estimate the angular width of the jet using equation 4 in Piran (2004).  Assuming a density of $n=1$ cm$^{-3}$, and an isotropic kinetic energy in the ejecta $\equiv \eta E_{\rm iso, \gamma}$, where $\eta$ is the radiative efficiency and we assume a value of $\eta=0.2$, we estimate an opening angle of $\theta\approx4.4^{\rm o}$.  Using $E_{\rm iso,\gamma}=2.3^{+0.3}_{-1.5} \times 10^{52}$, this in turn this implies a beaming-corrected $\gamma$-ray energy release of $E_{\rm \theta \gamma}=(\frac{\theta^{2}}{2})E_{\rm iso, \gamma} \sim 6.8\times 10^{49}$ erg.  If the density is higher, $n=10$, the opening angle is larger ($\theta\approx5.7^{\rm o}$), and so is the beam-corrected kinetic energy ($E_{\rm \theta, \gamma} \sim 1.2 \times 10^{50}$ erg).

\subsection{The Spectral Energy Distribution}
\label{sec:120729A_SED}

\begin{figure}
 \centering
 \includegraphics[bb=0 0 792 612, scale=0.31, clip=true,trim=0pt 0pt 0pt 0pt]{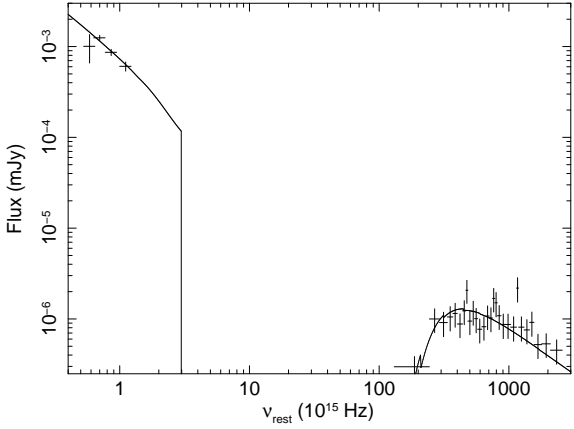}
 \caption{GRB 120729A: Rest-frame X-ray to optical SED of the AG at $t-t_{0}=0.42$ d.  It is found that a single PL provides a good fit to the data, with $\beta=1.0\pm0.1$.  Our best-fitting parameters ($\chi^{2}/dof=29.6/28$) are $A_{V}=0.15$ mag ($<0.55$ at 90\% CL), and an intrinsic column absorption of $N_{\rm H}=1.0\times10^{21}$ cm$^{-2}$ ($<2.7\times10^{21}$ at 90\% CL).  }
 \label{fig:120729A_AG_SED}
\end{figure}

We combined our host-subtracted GTC magnitudes at $t-t_{0}=0.75$ d, which were corrected for foreground extinction and converted into monochromatic fluxes, with contemporaneous X-ray observations to construct an X-ray to optical spectral energy distribution (SED), with the intention of getting an estimate of the amount of rest-frame extinction (Fig. \ref{fig:120729A_AG_SED}).  We used the general procedure outlined in Guidorzi et al. (2009) when constructing the energy spectrum.  As there are fewer X-ray photons at late times (the final observation is at $t-t_{0}=0.5$ d), the X-ray spectrum was accumulated from 0.046 to 0.074 days with 2.5 ks exposure.

Both a single ($\beta_{\rm X}=\beta_{\rm O}$) and broken PL ($\beta_{\rm X}-\beta_{\rm O}=0.5$, which is fixed) were fit to the SED, and it was found that a cooling break was not needed to fit the data, with a spectral index of $\beta=1.0\pm0.1$ proving to be a good fit.  When a cooling break was imposed upon the data, it was always found to occur below the optical data.  The paucity of data does not allow us to discriminate between the different extinction curves of the Small Magellanic Cloud (SMC), Large Magellanic Cloud (LMC) and Milky Way (MW) from Pei (1992), so we adopted the SMC template (which has proved to be a suitable fit to the AG SEDs, e.g. Kann et al. 2006).  Our best-fitting parameters ($\chi^{2}/\rm dof=29.6/28$) are $A_{V}=0.15$ mag ($<0.55$ at 90\% CL), and an intrinsic column absorption of $N_{\rm H}=1.0\times10^{21}$ cm$^{-2}$ ($<2.7\times10^{21}$ at 90\% CL).  To convert the rest-frame extinction into equivalent observer-frame extinctions in our SDSS filters we used the SMC extinction template at $z=0.8$ and the effective wavelengths in Fukugita et al. (1995), finding: $A_{g,\rm obs}=0.34$ mag, $A_{r,\rm obs}=0.26$ mag and $A_{i,\rm obs}=0.20$ mag.  We use these values of the rest-frame extinction throughout our analysis of GRB~120729A.

\subsection{The Host Galaxy}

\begin{figure}
 \centering
 \includegraphics[bb=0 0 864 504, scale=0.3]{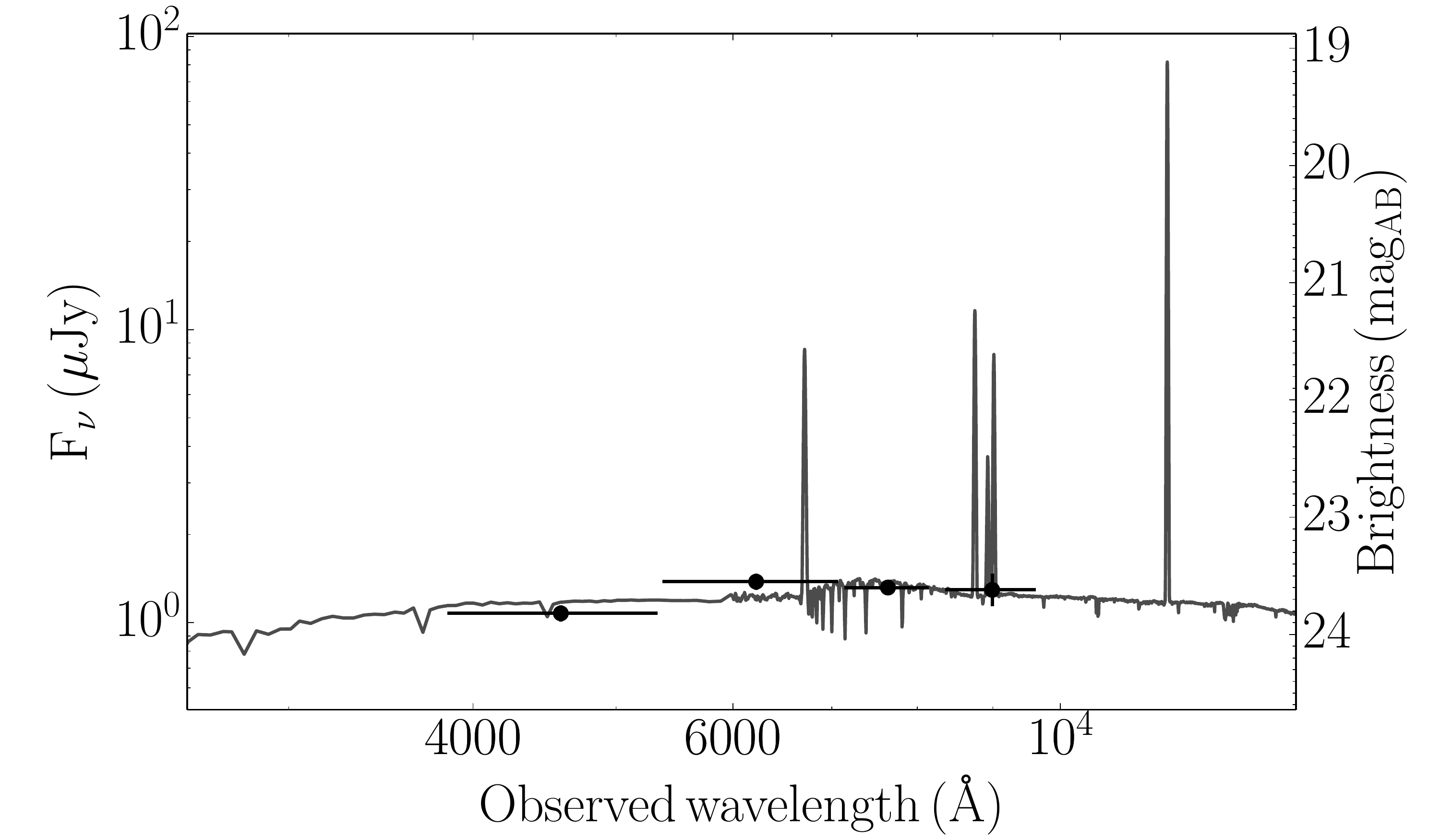}
 \caption{GRB 120729A: Best-fitting SED of the host galaxy $griz$ magnitudes.  The best fitting template is for a low-mass (log$_{10}$(mass)=$8.3\pm0.2$ $\rm M_{\odot}$), blue, star-forming galaxy (SFR$=6^{+25}_{-4}$ $\rm M_{\odot}$ yr$^{-1}$ and the age of the starburst $\le 100$ Myr).}
 \label{fig:120729A_host}
\end{figure}

We used our $griz$ observations of the host galaxy, taken at $t-t_{0}\approx189$ d and corrected for foreground extinction, to constrain some of its key physical properties (Fig. \ref{fig:120729A_host}).  Our procedure involves fitting the photometry with stellar population synthesis models from Bruzual \& Charlot (2003) with LePHARE (Arnouts et al. 1999).  We use a Calzetti dust attenuation law (Calzetti et al. 2000), a Chabrier initial mass function (Chabrier, 2003) and a grid of different star-burst ages with varying $e$-folding timescales to derive theoretical galaxy spectra which then were compared to our photometry. A more elaborate description of our SED fitting procedure and its caveats is given in Kr\"uhler et al. (2011).

The best fitting template is for that of a low-mass, blue, young star-forming galaxy.  The best-fitting parameters are: $\rm M_{B}=-19.3\pm0.1$, log$_{10}$(stellar mass)=$8.3\pm0.2$ $\rm M_{\odot}$, SFR$=6^{+25}_{-4}$ $\rm M_{\odot}$ yr$^{-1}$ and the age of the starburst $\le 100$ Myr.  The SFR has a large uncertainty due to the unknown dust attenuation.

\subsection{The Supernova}
\label{sec:120729A_SN}

The dearth of late-time observations limits our analysis of the accompanying SN to GRB~120729A, where only a few detections have been made near peak in $r$ and $i$.  Indeed the shape of the SN LC is not well constrained, especially given the lack of detections after the peak.  Nevertheless, despite this limitation we estimated the brightness of the SN in both filters during our fit.  In addition to fitting broken power-laws to LCs we included an additional SN-component.  Using the same C-code that was developed and adopted in C13\footnote{see as well Cano (2014, in prep.)}, we generated synthetic, k-corrected LCs of a template LC (SN~1998bw) as it would appear if it occurred at $z=0.80$.  Then using Pyxplot\footnote{http://pyxplot.org.uk} we fit the synthetic LC with a linear spline.  This spline is then incorporated into another function (equation 5 in C13) that transforms it by a stretch ($s$) and luminosity ($k$) factor.  In events where there are many observations of the SN bump (e.g. GRB~130427A / SN~2013cq, Xu et al. 2013a) it is possible to constrain both $s$ and $k$, however here we fix $s=1.0$ and allowed only $k$ to be a free parameter during the fit.  Our best-fitting parameters are: $k_{r}=1.29\pm0.19$ and $k_{i}=0.76\pm0.11$.  Taking these at face value implies (observer-frame) peak absolute magnitudes of $M_{r,\rm peak}=-18.96\pm0.15$ and $M_{i,\rm peak}=-19.29\pm0.15$, though these are somewhat tentative at best due to the uncertain stretch-factor of the SN, where a larger stretch factor implies a brighter luminosity.  With the same reasoning we have not attempted to estimate the time of peak light in each filter.

Using the method presented in C13, we estimated the bolometric properties of the accompanying SN.  Given that we have not been able to constrain the shape (i.e. width) of the SN in either filter, there is little merit in estimating its ejecta mass and in turn its kinetic energy.  However we can make an estimate of the amount of nickel that was nucleosynthesized during the explosion using ``Arnett's Rule'' (Arnett 1982) -- i.e. the luminosity at maximum is proportional to the instantaneous energy deposition from the radioactive decay of nickel and cobalt.  Making the assumption that the average luminosity factor of the accompanying SN in the optical filters is a suitable proxy for the relative difference in luminosity between this SN and the template, (which was shown in C13 to have an uncertainty of order 10\%), and using an average luminosity factor of $\bar{k}=1.02\pm0.26$, and fixing $s=1.0$, we estimate that in the filter range $UBVRIJH$ the accompanying SN has a nickel mass of $M_{\rm Ni}=0.42\pm0.11$ $M_{\odot}$.  The quoted error is statistical only, and arises from the uncertainty in the luminosity factor.  This nickel mass is close to that estimated for the archetypal GRB-SN~1998bw, where it is estimated 0.4--0.7 $M_{\odot}$ was nucleosynthesized (Iwamoto et al. 1998; Nakamura et al. 2001; Cano 2013).

\section{GRB 130215A}
\label{sec:130215A}

GRB 130215A was detected at 01:31:30 UT on 15-February-2013 by the \emph{Swift}-BAT, and has a $T_{90}= 65.7\pm10.8$ s in the 15--350 keV energy range (D'Elia et al. 2013; Barthelmy et al. 2013a).  Due to a Moon observing constraint, \emph{Swift} could not slew to the BAT position, thus there are no XRT or UVOT data for this GRB.   The burst was also observed by \emph{Fermi}-GBM (Younes \& Bhat, 2013) with $T_{90}\approx140$ s in the 50--300 keV energy range; and by the Sukaku Wide-Band All-sky Monitor (WAM) with $T_{90}\approx46$ s in the 100--1000 keV energy range (Ishida et al. 2013).  Rapid follow-up observations were performed by many ground-based telescopes (Zheng et al. 2013a, 2013b; LaCluyze et al. 2013; Cenko, 2013; Covino et al. 2013; Butler et al. 2013a, 2013b; Gendre et al. 2013; Xu \& Zhang, 2013; Hentunen et al. 2013a; Zhao \& Bai, 2013; Wren et al. 2013; Kuroda et al. 2013; Knust et al. 2013; Perley, 2013a, 2013b; Singer et al. 2013).  The redshift was measured to be $z=0.597$ (Cucchiara et al. 2013).  The AG was clearly detected at 93 GHz at $+2.73$ hr (Perley \& Keating 2013).  A spectrum of SN~2013ez was obtained with the GTC (de Ugarte Postigo et al. 2013a, 2013b).  An estimate of the isotropic energy release in $\gamma$-rays ($1-10^{4}$ keV, rest-frame) is $E_{\rm iso,\gamma}=3.1^{+0.9}_{-1.6} \times 10^{52}$ erg\footnote{http://butler.lab.asu.edu/swift/bat$\_$spec$\_$table.html}, and $E_{\rm p} = 155 \pm 63$ keV (Younes \& Bhat 2013).  The probability that GRB~130215A arises from a collapsar (Bromberg et al. 2013) based on $T_{90}$ alone is $99.995 \pm 0.002\%$ (BAT) and $99.487\pm0.358\%$ (GBM). We use a foreground extinction value of $E(B-V)_{\rm fore}$ = $0.162$ mag for GRB~130215A.

\subsection{Data Reduction, Photometry \& Spectroscopy}

We obtained observations with several ground-based telescopes.  ROTSE-III automatically started imaging the field of GRB~130215A 697 s after the initial $\gamma$-ray trigger, locating a new, bright (unfiltered=14.2) source at 02:54:00.73 +13:23:43.7 (J2000), with an uncertainty of $<1''$.  Further observations were obtained during the first day with the Nordic Optical Telescope (NOT) and the Gamma-Ray burst Optical/Near-Infrared Detector (GROND; Greiner et al. 2008).  Several hours of observations were obtained with the Reionization and Transients Infrared Camera (RATIR\footnote{www.ratir.org}) on the 1.5-m Harold Johnson Telescope at the Observatorio Astron\'omico Nacional on Sierra San Pedro M\'artir during the first few hours after the trigger, with additional epochs at $t-t_{0}=$2,3,4,8,11,17 d.  We also obtained two epochs of spectroscopy and one epoch of optical photometry with the GTC.  Early spectroscopy of the AG was performed with OSIRIS, $t-t_{0}=0.79$ d after the GRB using the R1000B grism, which gives a spectral resolution of $\delta\lambda/\lambda\sim1000$ and a coverage from 3600 to 7500 \AA.  We obtained an additional spectrum of the accompanying SN~2013ez at $t-t_{0}= 25.78$ d, the timing of which was planned to observe the SN at or near maximum light. This observation was performed using the R500R grism, with a spectral resolution of $\delta\lambda/\lambda\sim600$ and coverage from 4800 to 10000 \AA.  Each spectrum was reduced using standard techniques with IRAF-based scripts.  Late time images ($t-t_{0}=372.8$ d) of the GRB field were obtained with the GTC in filters $gri$, while a late epoch ($t-t_{0}=331.8$ d) was obtained with the 3.5-m CAHA telescope in $J$.

The optical data were calibrated via standard star photometry.  On 21-August-2013 GROND obtained images of the GRB field and an SDSS (Abazajian et al. 2009) field located at 03:00:48.0, +19:57:00 (J2000), with the SDSS field taken immediately after the GRB field.  Both sets of images were taken under photometric conditions.  The calibration was performed using a zeropoint and an airmass correction, and the solution was used to calibrate a set of secondary standards in the field of the GRB.  Each datatset was then calibrated to a subset of these stars, depending which ones were in the field of view of each telescope.  A summary of our photometry is presented in Table \ref{table:photometry}.

%

\subsection{The Afterglow}

\begin{figure}
 \centering
 \includegraphics[bb=0 0 263 189, scale=0.94]{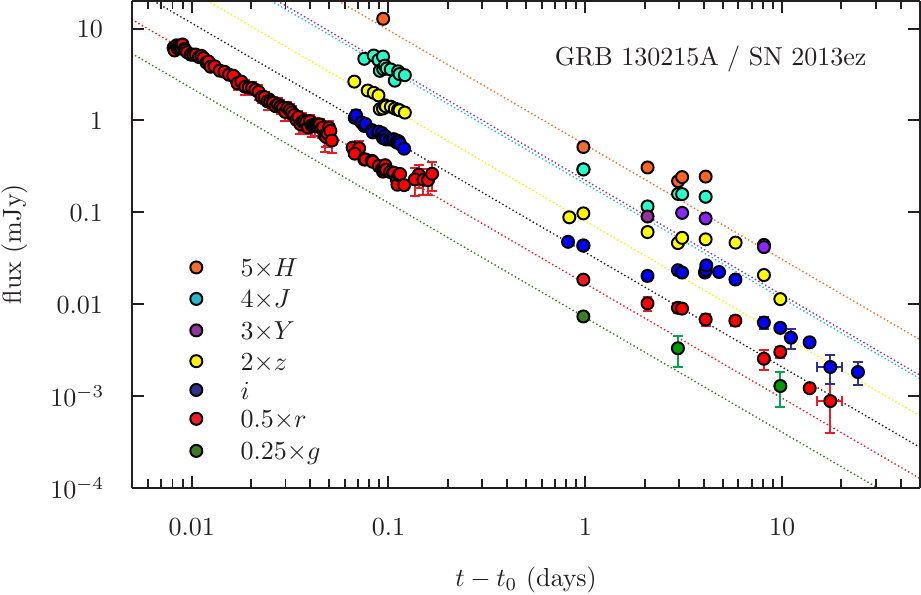}
 \caption{GRB 130215A: Optical and NIR LCs.  The optical data have been corrected for foreground extinction and converted into mJy using the flux zeropoints from Fukugita et al. (1995).  All filters have been fit with a single PL up to $t-t_{0}=1.0$ d (except Y which was normalised using the detection at 2.5 d), where $\alpha=1.25\pm0.01$ ($\chi^{2}/dof=231.9/141$).   A plateau is seen in all filters from $t-t_{0}=$1--6 d, where each LC deviates away from a single PL-like decline.}
 \label{fig:130215A_LC}
\end{figure}

Figure \ref{fig:130215A_LC} displays our optical photometry, which has been corrected for foreground extinction and then converted into monochromatic fluxes.  The LCs were simultaneously fit with a single PL up to $t-t_{0}=1.0$ d (except the Y-band data, which were normalised using the detection at 2.5 d, and is likely overestimated in brightness as this detection appears to be during the plateau phase), with the best-fitting value of the temporal index being $\alpha=1.25\pm0.11$ ($\chi^{2}/dof=231.9/141$).  After one day the LCs deviate away from the PL-like decline and undergo a plateau phase that lasts up to six days post burst.  The LCs then break again before leveling out a further time due to light coming from SN~2013ez.  


Due to the lack of XRT data we have not been able to construct an X-ray to optical energy spectrum.  Instead we used our contemporaneous optical/NIR data taken with RATIR and GROND over several epochs to get an estimate of the rest-frame extinction.  Using the same epochs as those used to construct the bolometric LC in Section \ref{sec:130215A_bolo_mag}, we fit the empirical extinction curves of the SMC, LMC and MW from Pei (1992), using a method similar to Kann et al. (2006) and Kann et al. (2010).  Each SED is well described by a single PL ($\beta=0.9\pm0.2$ over all epochs), with very little if any curvature, implying there is no need to invoke the presence of rest-frame dust extinction.  Each epoch is equally well fit by each dust-extinction template.  Indeed some epochs predict a (very small) negative value for the extinction, which is an unphysical conclusion, while the other epochs are consistent with zero rest-frame extinction.  Throughout the rest of the analysis of GRB~130215A, we assume $E(B-V)_{\rm rest}=0.0$ mag. 

\subsection{Magnetar Origins?}
\label{sec:130215A_bolo_mag}

\begin{figure}
 \centering
 \includegraphics[bb=0 0 263 189, scale=0.91]{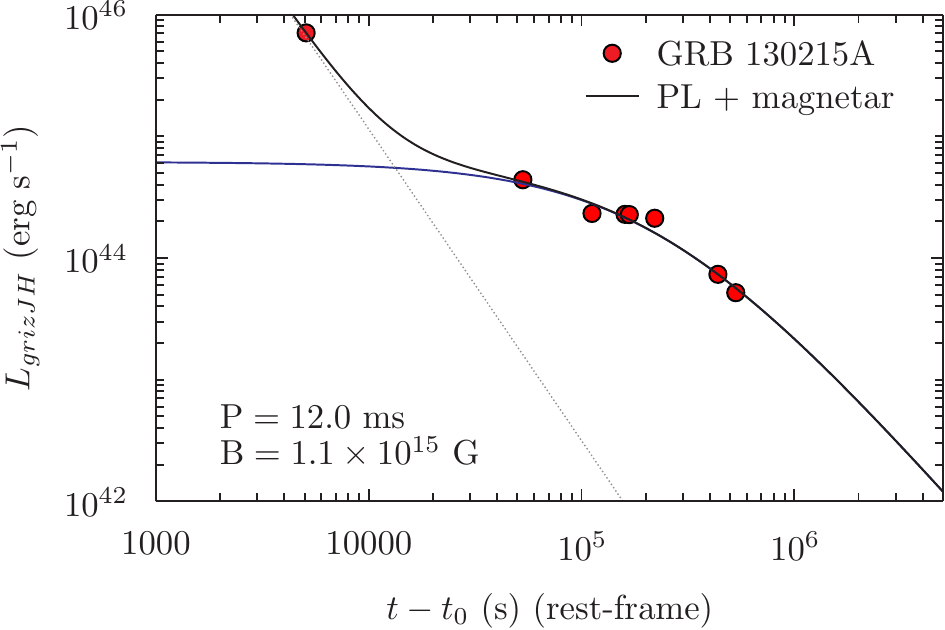}
 \caption{GRB 130215A: Rest-frame bolometric LC created from our $grizJH$ observations.  The analytical model from Zhang \& M\'esz\'aros (2001; see also Rowlinson et al. 2013) has been fit to the LC, which considers energy injection from a millisecond magnetar (plateau and late decline) added to an initial PL-like decline.  From the model we find an initial spin period of $\rm P_{0}=12.0$ ms, a magnetic field strength of $\rm B=1.1 \times 10^{15}$ G, a plateau luminosity of $\rm L_{plat}=6.1 \times 10^{44}$ erg s$^{-1}$ and a rest-frame plateau duration of $\rm T_{plat}=2.3 \times 10^{5}$ s (2.7d).  Encouragingly, the values of the initial spin period and $B$-field are \textit{realistic}, and commensurate to those found for long-duration GRBs 060729 and 130427A (see the main text), as well as the sample of short-duration GRBs in Rowlinson et al. (2013). }
 \label{fig:130215A_magnetar}
\end{figure}

There are many examples of GRB LCs that show deviations away from a PL-like decay, e.g. GRB~011211 (Jakobsson et al. 2003), GRB~021004 (de Ugarte Postigo et al. 2005), GRB~030429 (Jakobsson et al. 2004), GRB~050502A (Guidorzi et al. 2005) GRB~060526 (Th\"one et al. 2010), GRB~090926A (Rau et al. 2010; Cenko et al. 2011) and GRB~100814A (de Pasquale et al. 2013; Nardini et al. 2014).  One is also reminded of the peculiar LC of GRB~030329 (Matheson et al. 2003; Lipkin et al. 2004) that displayed very complex behaviour and complicated the decomposition of the SN light from the LC.  So while the AG LCs of some GRB-SNe are rather smooth (e.g. GRB~090618, Cano et al. 2011a), others are very complex.

The term ``energy injection'' is used to explain these peculiar bumps, flares and plateaus, where extra energy is pumped into the FS, causing the AG to become brighter (e.g. Panaitescu et al. 1998; Rees \& M\'esz\'aros 1998; Kumar \& Piran 2000; Sari \& M\'esz\'aros 2000).  Energy injection can arise from different physical sources including Poynting flux emitted by a central engine (e.g. Usov 1992; Dai \& Lu 1998), the collision of additional shells of material that collide with the original shells that generated the initial $\gamma$-ray burst (Zhang et al. 2006); a reverse shock (RS) created from the collision and pile up of multiple shells with the original shells (e.g. Sari \& M\'esz\'aros 2000; Kobayashi 2000; Harrison \& Kobayashi 2013; Japelj et al. 2014); a two-component jet (Granot et al. 2005) where a rebrightening in the optical bands can arise due to emission from a narrow jet seen off-axis; or a combination of forward and reverse shocks (de Pasquale et al. 2013) including the ``thick-shell'' scenario, where a combination of the forward and reverse shock (the latter is relativistic) leads to a plateau phase in the observations (Leventis et al. 2014) .  A more exotic source of energy injection can arise from a quark nova (Staff et al. 2008).  During the transition of the newly formed compact object from neutron star$\rightarrow$quark star$\rightarrow$black hole, accretion onto the quark star produces a source of extra energy that can be pumped into the ejecta, which can account for the prompt emission as well as flares and plateaus in X-ray LCs.  However, injection from an accreting quark star cannot explain plateaus in optical/NIR LCs.  One key idea that all these models have in common is that the later the energy injection episode, the more energy is required to create bumps and plateaus of similar magnitude.

Another source of energy injection into the FS can arise from a millisecond magnetar central engine, which deposits Poynting flux dominated dipole radiation into the ejecta (e.g. Zhang \& M\'esz\'aros 2001; Dall'Osso et al. 2011).  The millisecond magnetar model has been considered as a plausible source of energy injection for GRBs, with some notable examples being GRB~000301C (Zhang \& M\'esz\'aros 2001), GRB~060729 (Xu et al. 2009; Dall'Osso et al. 2011; L\"u \& Zhang 2014), GRB~120326A (Hou et al. 2014) and GRB~130427A (Bernardini et al. 2014).  In these investigations plateau phases in the X-ray LCs are attributed to extra energy arising from a millisecond magnetar, where energy injection refreshes the FS.  This is in contrast to the analysis of GRB~070110 (Troja et al. 2007), XRF 100316D (Fan et al. 2011), a sample of long-duration GRBs investigated by Lyons et al. (2010) and the recent study of a sample of short GRBs by Rowlinson et al. (2013), where both authors attribute the plateaus in the X-ray LCs as flux coming \textit{directly} from the millisecond magnetar.

To our knowledge, to date no attempt has been made to constrain the behaviour of a possible magnetar central engine using a bolometric LC of the AG constructed from optical/NIR observations.  Predominantly bolometric X-ray modelling has been the status quo, though an estimate of the optical contribution (\textit{R}-band) was made for GRB~130427A by Bernardini et al. (2014) and \textit{R}-band data of GRB~120326A (Hou et al. 2014).  In this work we are able to fully exploit the wide filter coverage and simultaneous observations obtained by both GROND and RATIR to create a bolometric LC in the filter range $grizJH$ (observer frame) with the aim of determining whether energy injection from a magnetar central engine provides a plausible explanation for the plateaus seen in the optical/NIR LCs.  We used data from a total of eight epochs (ranging from $t-t_{0}=0.1$--9.8 d in the observer frame); i.e. before the period where the SN starts to dominate the LCs).  We then followed a standard method to construct our bolometric LC (e.g. Cano et al. 2014): (1) correct all magnitudes for foreground and rest-frame extinction, (2) convert magnitudes into monochromatic fluxes using flux zeropoints in Fukugita et al. (1995).  For epochs where there are no contemporaneous observations, we linearly interpolated the flux LCs and SEDs to estimate the missing flux.  Then, for each epoch of multi-band observations, and using the effective wavelengths from Fukugita et al. (1995) we: (3) interpolate (linearly) between each datapoint, then (4) integrate the SED over frequency, assuming zero flux at the integration limits, and finally (5) correct for filter overlap.  The linear interpolation and integration were performed using a program written in Pyxplot.  The resultant LC is shown in Figure \ref{fig:130215A_magnetar}.  

We make similar assumptions as Rowlinson et al. (2013), namely that the magnetar mass is $1.4$ $M_{\odot}$ and its radius is $10^{6}$ cm, which allows us to reduce the number of free parameters in the fit.  The final fit is a combination of an initial PL added to the magnetar model:

\begin{equation}
L_{magnetar}(t)=L_{0}\left(1+\frac{t}{T_{0}}\right)^{-2} + \Lambda t^{-\alpha}
\label{equ:mag}
\end{equation}

\noindent where $L_{0}$ is the plateau luminosity, $T_{0}$ is the plateau duration, and $\Lambda$ is the normalisation constant for the PL.  The values of $L_{0}$ and $T_{0}$ can be related back to equations 6 and 8 in Zhang \& M\'esz\'aros (2001; see as well Rowlinson et al. 2013) to estimate the initial spin period and magnetic field strength of the magnetar.  Fitting this model to our rest-frame bolometric LC, we find an initial spin period of $\rm P_{0}=12.0$ ms, a magnetic field strength of $\rm B=1.1 \times 10^{15}$ G, a plateau luminosity of $\rm L_{0}=6.1 \times 10^{44}$ erg s$^{-1}$, a rest-frame plateau duration of $\rm T_{plat}=2.3 \times 10^{5}$ s, and $\alpha=2.6\pm0.7$.  

Encouragingly, the values of the initial spin and magnetic field are \textit{realistic}, and are found to be comparable to those found for other GRBs with associated SNe: (1) GRB~060729: $P=1.5$ ms and $B=0.27 \times 10^{15}$ G (Xu et al. 2009); $P=2.0$ ms and $B=3.2 \times 10^{15}$ G (Dall'Osso et al. 2011); and  $P=1.5$ ms and $B=0.25 \times 10^{15}$ G (Lu \& Zhang 2014); (2) GRB~130427A: $P\sim20$ ms and $B\sim 10^{16}$ G (Bernardini et al. 2014). The spin period determined from observations of GRB~130215A falls within the estimates for GRBs 060729 and 130427A, while the magnetic field strengths vary by two orders of magnitude for these three events.  Moreover, these values are fully consistent with the values determined for a sample of long GRBs by Lyons et al. (2010; their Table 2), a sample of short GRBs by Rowlinson et al. (2013; their Table 3), and for short GRB~130603B (de Ugarte Postigo et al. 2014).  Further discussion on the plausibility of energy injection arising from a millisecond magnetar is presented in Section \ref{sec:Discussion}.

\subsection{The Host Galaxy}
\label{sec:130215A_host}

We re-observed the field of GRB~130215A on 22-February-2014 with the GTC telescope in filters $gri$.  The host galaxy is not visible in any of our co-added images.  We derive 3-$\sigma$ upper limits for an isolated point source in our images of: $g>26.2$, $r>26.1$, $i>25.1$, which are not corrected for foreground extinction.  We also obtained a late-time J-band image with the 3.5-m CAHA telescope on 12-January-2014, where again no object is detected at the position of the GRB.  We derive an upper limit of $J>23.2$.

At $z=0.597$, and a distance modulus of $\mu=42.72$, these upper limits imply observer-frame, absolute magnitude limits of the host galaxy of $M_{g}>-16.5$, $M_{r}>-16.6$, $M_{i}>-17.6$ and $M_{J}>-19.5$.

\subsection{The Supernova}

\subsubsection{Spectroscopy of SN 2013ez}

\begin{figure}
 \centering
 \includegraphics[bb=0 0 576 432, scale=0.45]{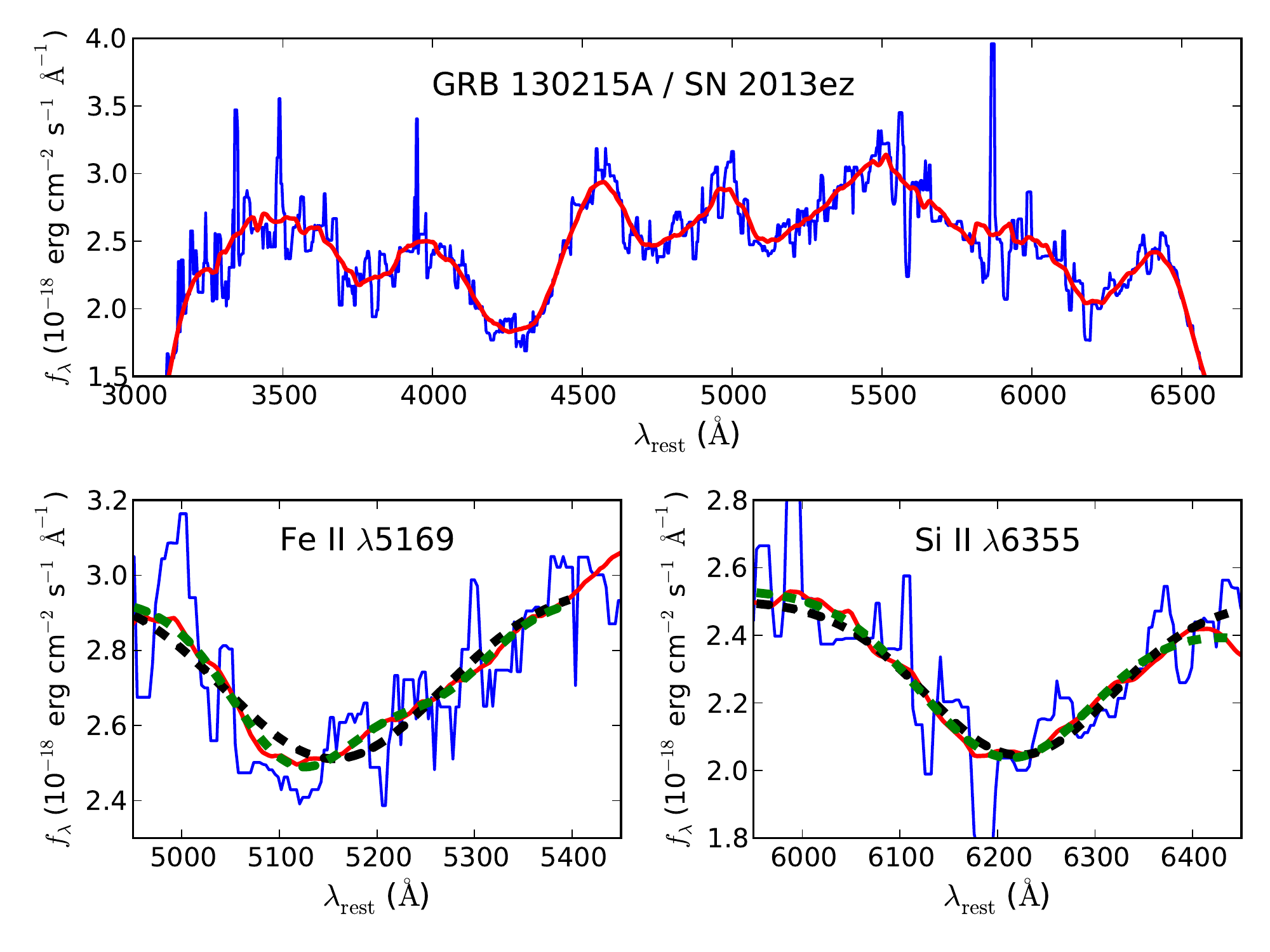}
  \caption{GRB 130215A / SN~2013ez: SN+host+AG (blue) rest-frame spectrum ($\lambda$ in air), from $t-t_{0}=$ 25.8 d (16.1 d rest-frame).  \textit{Top:} The spectrum in blue has been median filtered (width of 20 \AA), while the red spectrum has been smoothed using a Kaiser window ($M=40$, $\beta=0$).  Clear undulations are seen in the spectrum that are reminiscent of other SNe Ic.  \textit{Bottom left:}  Close-up of the absorption feature near 5100 \AA $ $, which is thought to be blueshifted FeII $\lambda$5169.  Both a single and double Gaussian have been fit to the smoothed spectrum.  The latter provides a better fit ($\lambda_{1}=5100.59\pm24.45$ \AA $ $, $\rm v_{Fe} = -4000\pm1496$ km s$^{-1}$; $\lambda_{2}=5305.72\pm202.17$ \AA $ $, which is too red to be \ion{Fe}{ii} $\lambda$5169).  \textit{Bottom right:}  Close-up of the absorption feature near 6200 \AA $ $, which is thought to be blueshifted \ion{Si}{ii} $\lambda$6355.  A single Gaussian was fit to the spectrum, yielding $\lambda_{1}=6221.75\pm16.77$ \AA $ $, $\rm v_{Fe} = -6354\pm808$ km s$^{-1}$.  A double Gaussian was also fit to the smoothed spectrum, however an improved fit was not obtained.  The blueshifted velocities of both lines are comparable within their respective error bars, and show that SN 2013ez may be more appropriately classified as a type Ic SN rather than a type Ic-BL.}
   \label{fig:130215A_spectrum}

\end{figure}

\begin{figure}
 \centering
 \includegraphics[bb=0 0 263 189, scale=0.99]{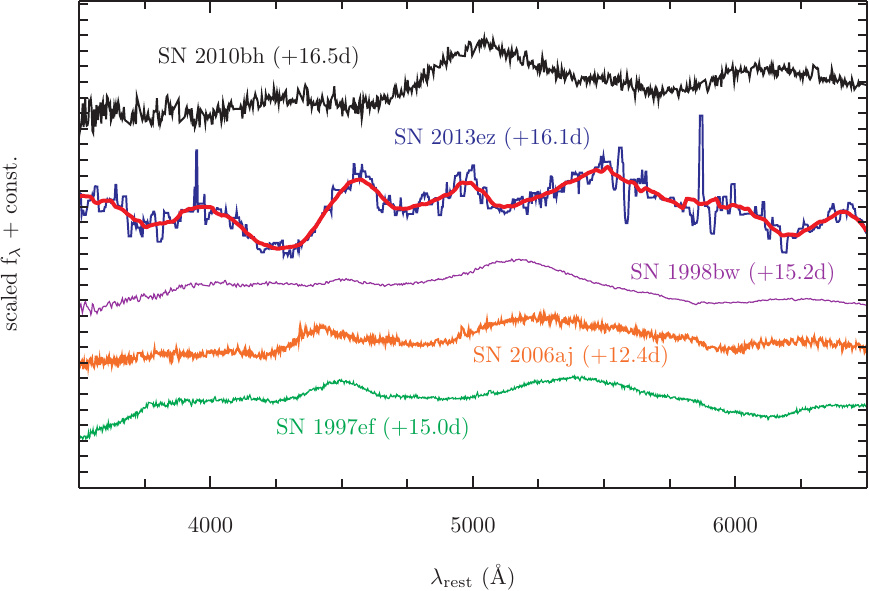}
 \caption{GRB 130215A / SN~2013ez: For comparison, we plot the spectra of several GRB-SNe (SN 1998bw, purple line; SN 2006aj, orange line; SN 2010bh, black line) and one SN IcBL (SN 1997ef, green line) that was not associated with a GRB, each at a similar post-explosion date as SN 2013ez.  All times are rest-frame and the spectra have been scaled and shifted for clarity.  It is seen that the absorption features in the spectrum of 2013ez are not as broad as those of the comparison SNe. }
 \label{fig:130215A_spectrum_IcBL}
\end{figure}

\begin{figure}
 \centering
 \includegraphics[bb=0 0 263 189, scale=0.99]{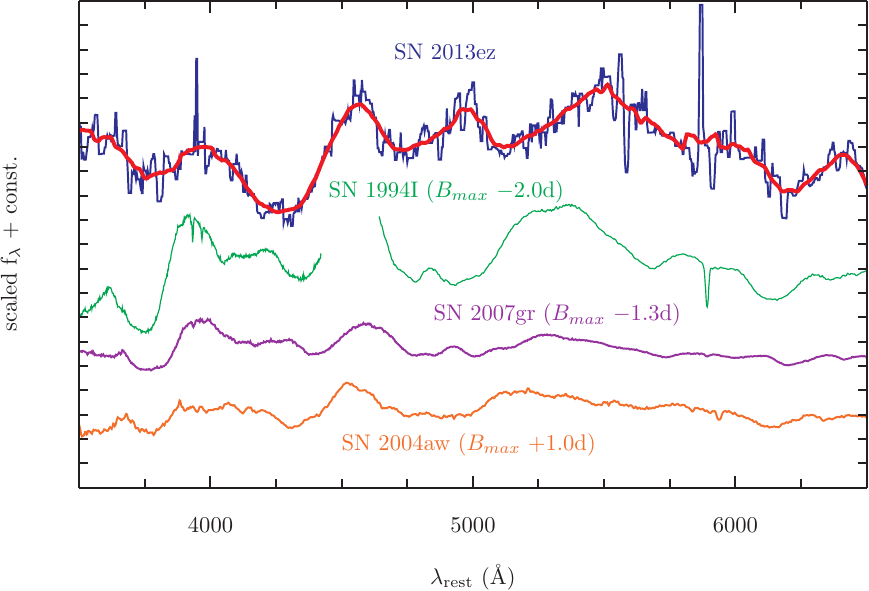}
 \caption{GRB 130215A / SN~2013ez: Plotted for comparison are the rest-frame spectra of several Ic SNe: 1994I (green), SN~2004aw (orange) and SN~2007gr (purple).  All times are from the time of maximum $B$-band light for each SN.  The narrow features of SN 2013ez are more reminiscent of those of SNe Ic.}
 \label{fig:130215A_spectrum_Ic}
\end{figure}

\begin{figure}
 \centering
 \includegraphics[bb=0 0 265 191, scale=0.95]{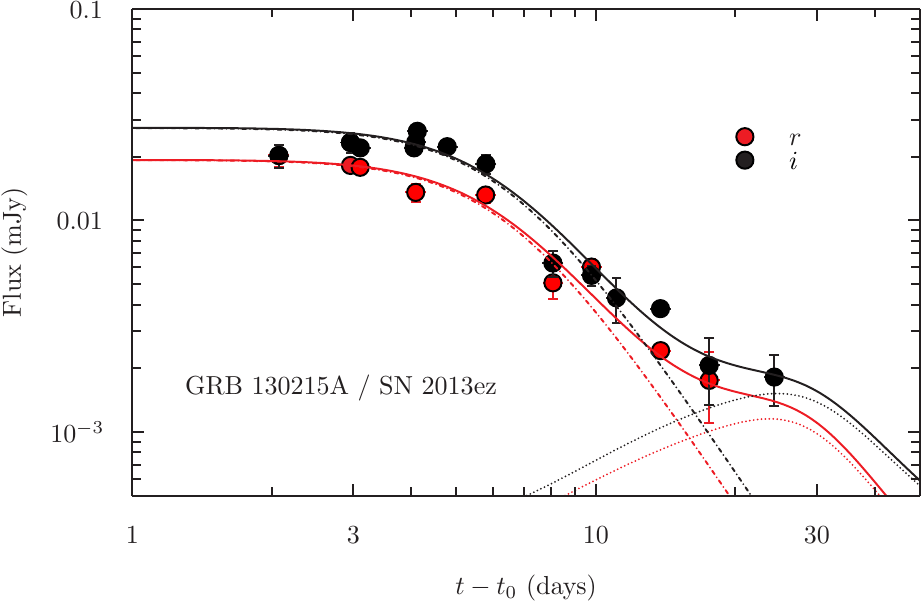}
 \caption{GRB 130215A / SN~2013ez: Optical LCs in $r$ (red) and $i$ (black).  The AG (dot-dashed) and SN (dotted) components are shown in the same colour as their corresponding filter, while the solid lines are the sum of both components.  Data at times $>+2$ d have been fit with a broken PL consisting of a plateau phase and a break to a steeper decay phase.  We assume that the time at which the LC breaks and the decay rate after the break are the same in both filters.  The best-fitting values (for magnitudes that are not host-subtracted, see the main text) are: $\alpha_{2}=3.28\pm0.25$ and $T_{B}=6.39\pm0.35$ d.  Due to the (1) lack of host detection in our deep GTC images, and (2) lack of datapoints at times when the SN is the dominant source of flux we have not been able to precisely constrain the SN's properties.  When we consider the two extremes of the host brightness (see main text), we constrain the luminosity factor of SN~2013ez to be $0.6\le k \le 0.75$.}
 \label{fig:130215A_SN}
\end{figure}

The rest-frame (wavelength) spectrum of SN 2013ez at $t-t_{0}=25.8$ d (16.1 d rest-frame) is displayed in Fig. \ref{fig:130215A_spectrum}.  The spectrum in blue is the result of applying a moving median filter (width=20 \AA) to the original spectrum.  In order to more clearly distinguish the key absorption features, we smoothed the spectrum using a Kaiser window as a smoothing kernel, where the size of the window ($M$) and shape of the window ($\beta$) were varied.  The smoothing procedure was performed using a Python program that utilizes numpy to convolve the spectrum by the Kaiser window.  In Fig. \ref{fig:130215A_spectrum} values of $M=40$ and $\beta=0$ are used.

Two absorption features are seen near 5100 \AA $ $ and 6200 \AA $ $, which are thought to be blueshifted \ion{Fe}{ii} $\lambda$5169 and \ion{Si}{ii} $\lambda$6355 respectively.  We fit a single and double Gaussian to both features in the smoothed spectrum in order to determine their central wavelengths and blueshifted velocities (see bottom insets in Fig. \ref{fig:130215A_spectrum}).  In order to get an estimate of the error of each wavelength (and velocity) measurement, we performed a Monte Carlo (MC) simulation to generate 10,000 spectra from the original spectrum and its error spectrum.  At each wavelength in the original spectrum, we derived a random number from a Gaussian distribution that is centered at each wavelength in the spectrum, and whose standard deviation is equal to the value of the error spectrum at that wavelength.  We also allowed $M$ and $\beta$ to be random numbers between two pre-determined values.  Then, in the MC simulation we fit both a single (SG) and double Gaussian (DG) to to the \ion{Fe}{ii} and \ion{Si}{ii} features, calculating their central wavelengths and blueshifted velocities.  Presented below are the average values of the 10,000 computed wavelengths and velocities, where the 1-$\sigma$ errors have been calculated from their respective standard deviations.  The errors are a combination of those arising from the error spectrum and the fitting procedure.  The systematic error in $\lambda$ and the blueshifted velocities arising from fitting the absorption features in a single spectrum 10,000 times are $\approx 2$ \AA, and $\approx$ 100 km s$^{-1}$ respectively.  The errors below are clearly dominated by the error spectrum.

The results of our MC simulation for \ion{Fe}{ii} $\lambda$5169 are: (SG): $\lambda=5125.27\pm24.90$ \AA; $v=-2552\pm1464$ km s$^{-1}$, (DG): $\lambda_{1}=5100.59\pm25.45$ \AA; $v_{1}=-4000\pm1496$ km s$^{-1}$; $\lambda_{2}=5305.72\pm202.17$ \AA.  $\lambda_{2}$ is clearly too red to be \ion{Fe}{ii} $\lambda$5169, and so we have not computed its blueshifted velocity.

Similarly, for \ion{Si}{ii} $\lambda$6355: (SG): $\lambda=6221.75\pm16.77$ \AA; $v=-6357\pm808$ km s$^{-1}$, (DG): $\lambda_{1}=6172.44\pm569.46$ \AA; $v_{1}=-8284\pm6219$ km s$^{-1}$; $\lambda_{2}=6156.27\pm7406.63$ \AA;  $v_{1}=-1643\pm11,544$ km s$^{-1}$.  Clearly the double Gaussian does not improve the fit, and the central wavelengths of both are approximately equal.

In summary, the \ion{Fe}{ii} $\lambda$5169 feature is better described by a DG, where a blueshifted velocity of $v\approx-4000\pm1500$ km s$^{-1}$ is found.  The \ion{Si}{ii} $\lambda$6355 is better described by a SG, where a blueshifted velocity of $v\approx-6350\pm800$ km s$^{-1}$ is found.  The velocities, of quite modest values, agree within their respective errorbars.  

To put these velocities into context, Schulze et al. (2014) investigated the velocity of \ion{Fe}{ii} $\rm \lambda5169$ for a sample of seven GRB-SNe.  At $t-t_{0}=$16 d (rest-frame), the velocity of this line ranges from $\sim-$14,000 km s$^{-1}$ for SN~2006aj to $\sim-$36,000 km s$^{-1}$ for SN~2010bh.  The velocity of this line in the spectrum of SN~2013ez is $\sim$8000--10,000 km s$^{-1}$ \textit{slower} than that of SN~2006aj, which itself represents the slowest velocity among all the GRB- and XRF-SNe in their sample.  Based on this argument it may be more appropriate to classify SN~2013ez as type Ic rather than Ic-BL.

Plotted for comparison\footnote{The majority of the comparison spectra were downloaded from the WiseREP SN spectrum database (Yaron \& Gal-Yam, 2012), with the rest coming (with our gratitude) from their respective references.} in Fig. \ref{fig:130215A_spectrum_IcBL} are the spectra of SN~1998bw (Patat et al. 2001) that was associated with GRB~980425, SN~2006aj (Modjaz et al. 2006) that was associated with XRF 060218, SN~2010bh (Bufano et al. 2012) that was associated with XRF 100316D, and SN~1997ef (Garnavich et al. 1997; Hu et al. 1997) which is a SN IcBL not associated with a GRB.  Each spectrum is taken near a similar rest-frame time to that of SN~2013ez: SN~1998bw ($t-t_{0}=$15.2 d; Patat et al. 2001), SN~2006aj ($t-t_{0}=$12.4 d; Modjaz et al. 2006), SN~2010bh ($t-t_{0}=$16. 5d; Bufano et al. 2011), and SN~1997ef ($t-t_{0}\approx$15 d; Mazzali et al. 2000).  The majority of the telluric and host galactic lines have been removed, and the spectra have been shifted and scaled to provide a clear comparison.  It is seen that the absorption features in the spectrum of SN~2013ez are not as broad as those of the comparison SNe.  

Conversely, plotted in Fig. \ref{fig:130215A_spectrum_Ic} are three Ic SNe not associated with a GRB near maximum $B$-band light: SN~1994I ($B_{max}-2$d ; Filippenko et al. 1995), SN~2004aw ($B_{max}$+1 d; Taubenburger et al. 2006) and SN~2007gr ($B_{max}-1.3$ d; Valenti et al. 2008).  The narrow lines in the comparison SNe Ic are redolent of those of SN~2013ez.

\subsubsection{Photometry of SN 2013ez}

While the spectroscopic identification of SN~2013ez is unambiguous, the plateau in the optical/NIR complicates our ambition of decomposing the LCs in order to isolate the photometric SN contribution.  The situation is also perplexed by the paltriness of photometric observations of SN~2013ez near peak.  Nevertheless we decomposed the optical LCs to estimate the brightness of SN~2013ez in filters $r$ and $i$ (Fig. \ref{fig:130215A_SN}).  We fit a broken PL to the LCs, and imposed a plateau phase ($\alpha_{1}=-0.01$), which breaks at some time ($T_{\rm B}$) to a steeper decay phase ($\alpha_{2}$).  We assume that the time the LC breaks and the rate of decay after the break are the same in both filters, and these two parameters are allowed to vary during the fit.

The decomposition is further complicated by the fact that we have not detected the host in our deep GTC images (see section \ref{sec:130215A_host}), so we must consider two scenarios: (1) magnitudes that are not host-subtracted, and (2) take the host brightness equal to the limits obtained from the GTC images (and correct for foreground extinction).  These two scenarios can be considered to be the two extremes to the SN brightness, for certainly the host will be contributing \textit{some} flux, but no more than the upper limits of the GTC images.

In scenario (1) we find best-fitting AG parameters of $\alpha_{2}=3.28\pm0.25$ and $T_{B}=6.39\pm0.35$ d, while in scenario (2) we find $\alpha_{2}=3.44\pm0.28$ and $T_{B}=6.41\pm0.34$ d.  Unsurprisingly the time the LC breaks is essentially the same in both scenarios, while the LCs decay faster when we remove a host contribution.  We also note that the break-time is later than that found in the magnetar model ($T_{\rm B}=4.3$ d in observer frame).  As for GRB~120729A, we fix the stretch factor to $s=1.0$ due to the lack of datapoints.  In scenario (1) we find $k\approx 0.75$, while in scenario (2) we find $k\approx0.6$.  These two values can be considered the upper and lower limits to the brightness of SN~2013ez in these filters.  Taking these values at face value implies peak brightnesses of $M_{r}=-18.7$ to $-19.0$, and $M_{i}=-19.0$ to $-19.3$.  Again, there is little merit in estimating the peak times due to the unknown stretch values.  Finally, using the method in C13, we estimate a nickel mass in the range $0.25 \le M_{\rm Ni} \le 0.30$ $M_{\odot}$.

\section{GRB 130831A}
\label{sec:130831A}

GRB 130831A was detected at 13:04:16 UT on 31-August-2013 by the \emph{Swift} Burst Alert Telescope (BAT), and has a $T_{90}= 32.5\pm2.5$ s in the 15--350 keV energy range (Hagen et al. 2013; Barthelmy et al. 2013b).  It was also detected by Konus-Wind,  Golenetskii et al. (2013) estimate an isotropic energy release in $\gamma$-rays of $E_{\rm iso,\gamma}=4.6\pm0.2\times 10^{51}$ erg in the 20 keV--10 MeV range, and $E_{\rm p} = 67 \pm 4$ keV (Golenetskii et al. 2013).  The probability that GRB~130831A arises from a collapsar (Bromberg et al. 2013) based on $T_{90}$ alone is $99.969 \pm 0.006\%$ (BAT).

Rapid follow-up of GRB~130831A was performed by several ground-based telescopes (Guidorzi \& Melandri 2013; Xu et al. 2013b; Yoshii et al. 2013; Xin et al. 2013; Trotter et al. 2013; Leonini et al. 2013; Masi \& Nocentini 2013; Izzo \& D'Avino 2013; Hentunen et al. 2013b; Sonbas et al. 2013; Butler et al. 2013c; Chester \& Hagen 2013; Volnova et al. 2013a, 2013b, 2013c, 2013d; Pozanenko et al. 2013 and Khorunzhev et al. 2013).  The AG was not detected at radio (Laskar et al. 2013) or sub-mm wavelengths (Zauderer et al. 2013; Smith et al. 2013).  A redshift of $z=0.479$ was measured with Gemini-North (Cucchiara \& Perley 2013).  A spectrum of the associated supernova, SN~2013fu, was obtained with the VLT by Klose et al. (2013).  We use a foreground extinction value of $E(B-V)_{\rm fore}$ = $0.046$ mag for GRB~130831A.

\subsection{Data Reduction \& Photometry}

We obtained optical observations with several ground-based telescopes.  The 0.65-m SANTEL-650 and 0.5-m VT-50 telescopes of the UAFO/ISON-Ussuriysk started imaging (unfiltered) the GRB field just over 10 minutes after the initial trigger, obtaining nearly consecutive images for six straight hours.  Additional follow-up observations were obtained with the Gissar observatory 0.7-m telescope, the 0.4-m SANTEL-400AN telescope (ASC/ISON-Kislovodsk observatory), the 0.7-m AZT-8 telescope operated by the Institute of Astronomy, Kharkiv National University and the 1.5-m AZT-33IK telescope at Mondy observatory, Shajn 2.6-m telescope of Crimean Astrophysical observatory and the 1.5-m AZT-22 telescope at Maidanak observatory.  Data obtained at times $t-t_{0}<2.0$ d with aforementioned Russian telescopes are presented in de Pasquale et al. (2014, in prep), while everything at this time and later are presented in this paper.  We obtained several epochs of photometry with the 2.5-m NOT, three epochs with the 4.2-m William Hershel telescope (WHT), and four epochs with the 2.0-m LT.  We also obtained a single late-time epoch with the Gemini Multi-Object Spectrograph (GMOS; Hook et al. 2004) mounted at Gemini-South as a part of the program GS-2013B-Q-69.   The data were reduced in a standard fashion with the Gemini IRAF software package for GMOS (v1.12).  

The optical data were calibrated using SDSS stars in the GRB field with a zeropoint between the instrumental and catalog magnitudes.  Observations obtained with each Russian telescope in Johnson/Cousins filters $BVR_{c}I_{c}$ were calibrated by converting the SDSS (AB) magnitudes of local standards into Johnson/Cousins (Vega) using transformation equations in Lupton (2005).   The late-time $r$ observations taken with the NOT, WHT and Gemini were converted into $R_{c}$ using transformation equations from Jordi et al. (2006), which require a colour term ($r-i$) in the calculations.  In an identical procedure as for GRB~120729A (see section \ref{sec:120729A_datared}) we interpolated the $i$-band LC to the times of the $r$-band LC, extracting the $i$ magnitude.  A summary of our photometry is presented in Table \ref{table:photometry}.

\subsection{The Afterglow}
\label{sec:130831A_AG}

Our $R_{c}iz$ optical data are displayed in Fig. \ref{fig:130831A_LC}.  All magnitudes are corrected for foreground and rest-frame extinction.  We host-subtracted the optical data in $R_{c}$ and $i$ using the host detections (see Section \ref{sec:130831A_host}) in the same filters, and then converting all magnitudes into monochromatic fluxes and then mathematically subtracting the host flux from the earlier epochs.  The $z$ data have not been host-subtracted due to lack of observations of the host in $z$ at late times.  All LCs are well described by a single PL, where we assume that the LCs decay at the same rate in all filters, where $\alpha=1.63\pm0.02$.  We note the presence of a bump or short plateau phase in the \textit{R}-band LC between $t-t_{0}\sim$ 3--5 d, however this short phase does not appear to affect our analysis of the decay rate and subsequent optical properties of the SN in $R_{c}$ and $i$.

\subsection{The Spectral Energy Distribution}
\label{sec:130831A_SED}

\begin{figure*}
 \centering
 \includegraphics[bb=0 0 536 1423, angle=270, scale=0.31]{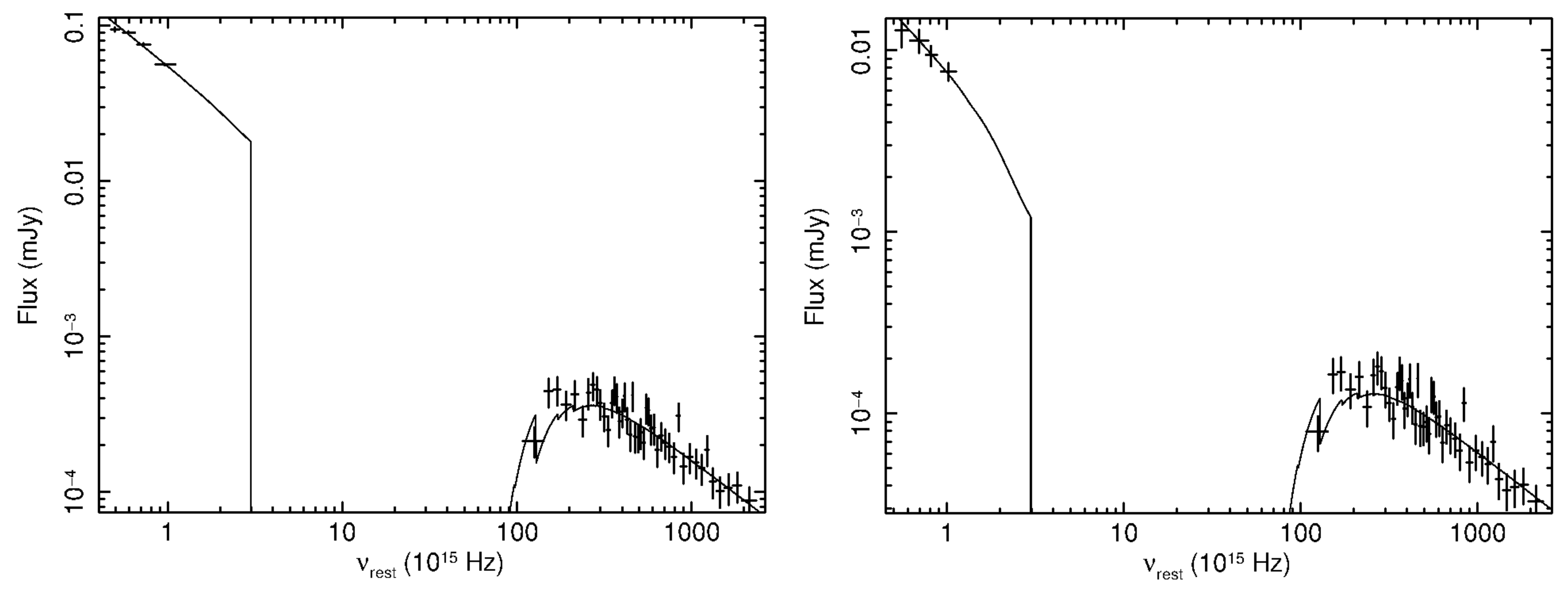}
 \caption{GRB 130831A: X-ray to optical SED of the AG at $t-t_{0}=0.39$ and 1.12 d (observer frame).  It is found that a single PL provides a good fit to both epochs of data ($\beta=0.85\pm0.01$ and $0.75\pm0.06$ respectively).  For the first epoch (left) we find  $A_{V}<0.1$ mag (90\% CL), while in the second epoch (right) we find $A_{V}=0.21^{+0.28}_{-0.21}$ mag.  The results of the SED fitting indicate that the rest-frame extinction is consistent with being $A_{V}=0.0$ mag.  Additionally we find the intrinsic column density to be $N_{\rm H}\approx 3-4 \times 10^{20}$ cm$^{-2}$ for both epochs.   }
 \label{fig:130831A_SED}
\end{figure*}

In an identical analysis as Section \ref{sec:120729A_SED} we constructed rest-frame X-ray to optical SEDs in order to get an estimate of the rest-frame extinction (Fig. \ref{fig:130831A_SED}).  We fit two epochs of data at $t-t_{0}=0.39$ d and 1.12 d (observer frame), which coincide with optical data obtained with the NOT and MAO respectively.  The X-ray spectra were derived from 0.11--0.61 d with a total exposure of 8 ks and rescaled at both epochs of optical data.  The optical data have been corrected for foreground extinction.

As before both single and broken PLs were fit to the SEDs, and we find that for both epochs a single PL fits the data well.  Our results for both epochs are: (1) $t-t_{0}=0.39$ d ($\chi^{2}/\rm dof=49.2/45$): $\beta=0.85\pm0.01$, $A_{V}<0.1$ mag (90\% CL), and $N_{\rm H}=4.2\pm0.8\times10^{20}$ cm$^{-2}$; (2) $t-t_{0}=1.12$ d ($\chi^{2}/dof=42.0/45$): $\beta=0.75\pm0.06$, $A_{V}=0.21^{+0.28}_{-0.21}$ mag, and an intrinsic column density of $N_{\rm H}=3.5\pm1.0\times10^{20}$ cm$^{-2}$.  We thus conclude that the extinction local to GRB~130831A is consistent with being zero, and for our analysis we use the value of $E(B-V)_{\rm rest}=0.0$ mag.

\subsection{The Host Galaxy}
\label{sec:130831A_host}

We observed the field of GRB~130831A at late times with the LT ($i$) and NOT ($r$), and find an extended object at the GRB position in both images, which we attribute as the host galaxy.  In the $i$-band LT image taken on 05-Jan-2014 at $t-t_{0}=127.3$ d ($+86.1$ d in rest-frame), we measure $i=24.23\pm0.10$.  In our $r$-band NOT image taken on 03-Feb-2014 at $t-t_{0}=156.3$ d ($+105.7$ d in rest-frame), we measure $r=24.06\pm0.09$.  These magnitudes are not corrected for foreground extinction.  In terms of absolute magnitude, we find (observer-frame) $M_{r}=-18.06\pm0.09$ and $M_{i}=-17.89\pm0.10$.  The colour $r-i=-0.17$ suggests that the host galaxy is rather blue.  We note that we did not attempt to fit galaxy SEDs due to the sparseness of host observations.

\subsection{The Supernova}

\begin{figure}
 \centering
 \includegraphics[bb=0 0 263 189, scale=0.93]{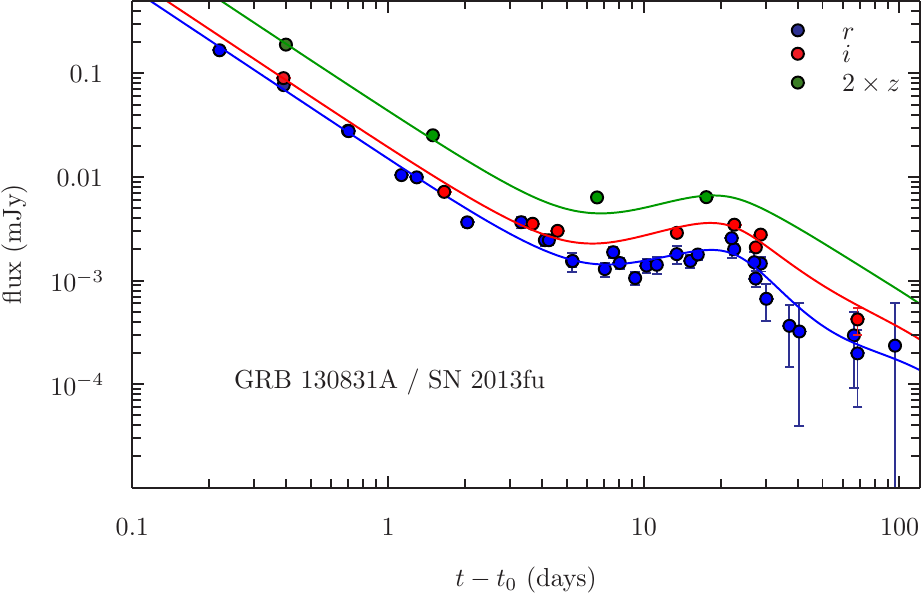}
 \caption{GRB 130831A: Optical ($R_{c}iz$) light curves.  The solid lines in each filter are the sum of the AG and SN components.  The optical data have been corrected for foreground extinction, and the $R_{c}$ and $i$ are host-subtracted (see the text).  The $z$ data have not been host-subtracted due to lack of observations of the host in $z$ at late times.  All LCs are well fit with a single PL, and we assume that the LCs decay at the same rate in all filters, where $\alpha=1.63\pm0.02$.  A clear SN bump is seen in $R_{c}$ and $i$, and a flattening of the LC is seen in $z$, which can be attributed to flux coming from SN~2013fu.  In each filter we simultaneously fit a SN-component to determine the stretch ($s$) and luminosity ($k$) factors in each filter.  Due to the lack of observations in $z$ at late times we fix the value of stretch factor to be the same in $i$ and $z$ (i.e. $s_{i} \equiv s_{z}$).  Our best-fitting parameters are: $k_{R}=0.65\pm0.03$ and $s_{R}=0.84\pm0.03$;  $k_{i}=1.04\pm0.05$ and $s_{i}=0.81\pm0.03$; $k_{z}=1.02\pm0.19$ and $s_{z}=s_{i}=0.81$ (fixed).  As the $z$ data are not host-subtracted, the luminosity factor is an upper limit to the maximum brightness of SN~2013fu in this filter.}
 \label{fig:130831A_LC}
\end{figure}

A clear SN bump in filters $riz$, which was initially suggested as the signature of SN~2013fu (Pozanenko et al. 2013b), is seen in Fig. \ref{fig:130831A_LC}.  The SN bump is particularly pronounced in the well-sampled \textit{R}-band LC, and is also seen in $i$, and to a lesser extent in $z$, where a flattening of the LC is seen, which can be attributed to flux coming from SN~2013fu.  When fitting the optical data in Section \ref{sec:130831A_AG} we simultaneously fit a SN-component to determine the stretch ($s$) and luminosity ($k$) factors in each filter.  Our best-fitting parameters are: $k_{R}=0.65\pm0.03$ and $s_{R}=0.84\pm0.03$ $k_{i}=1.04\pm0.05$ and $s_{i}=0.81\pm0.03$; $k_{z}=1.02\pm0.19$ and $s_{z}=s_{i}=0.81$ (fixed).  Due to the lack of observations in $z$ at late times we fix the value of $s$ to be the same as in $i$.  Moreover, as the $z$ data are not host-subtracted, the luminosity factor is an upper limit to the maximum brightness of SN~2013fu in this filter.

We determined the peak absolute magnitude, and time of peak light, of SN~2013fu in each filter.  In $R_{c}$: $M_{R}=-18.89\pm0.05$ and $t_{p}=18.60\pm0.67$ d ($12.58\pm0.45$ d in rest frame); $i$: $M_{i}=-19.56\pm0.05$ and $t_{p}=18.53\pm0.069$ d ($12.53\pm0.47$ d in rest frame); $z$: $M_{z}=-19.47\pm0.19$ and $t_{p}\approx19.1$ d ($\approx12.9$ d in rest frame).  The peak time in $z$ is tentative however, given that we have not been able to directly determine the stretch factor directly from our observations.  At $z=0.479$, observer-frame $i$ ($\lambda_{\rm eff}=7706$ \AA) is roughly rest-frame $V$ ($\lambda_{\rm eff}=5505$ \AA): $7706/1.4791=5210$ \AA.  Making a k-correction using the formulation of Hogg et al. (2002), and a spectrum of SN~1998bw as a template, we find a k-correction from observer-frame $i$ to rest-frame $V$ of k$_{i\rightarrow V}\approx0.25$ mag.  This implies a rest-frame, peak magnitude of $M_{V}\approx-19.31$.  This value is consistent with the average peak V-band magnitude found for a sample of k-corrected LCs of GRB-SNe analysed by Richardson (2009), who found $M_{V\rm,peak}=-19.2\pm0.2$ (standard deviation of $\sigma=0.7$ mag).

It is seen that in the well-sampled \textit{R}-band LC that the SN appears to decrease in brightness faster than the k-corrected LC of SN~1998bw.  There is also a hint of this in the \textit{I}-band LC, though we cannot draw many conclusions based on a single datapoint at late times.  When we calculate $\Delta m_{15}$ in $R_{c}$ (where $\Delta m_{15}$ is the amount the LC fades from peak light to 15 days later; which is calculated for rest-frame times) for SN~2013fu and SN~1998bw, (where the latter is transformed by $k_{R}=0.65\pm0.03$ and $s_{R}=0.82\pm0.03$), we find $\Delta m_{15}\approx+1.99$ and $\Delta m_{15}\approx+1.45$ respectively.  This clearly shows that 2013fu evolves faster than the archetype GRB-SN~1998bw.  Therefore, in this case the shape of the template SN does not provide the best description for the temporal evolution of SN~2013fu.  This type of behaviour has been seen for other GRB-SNe, such as SN~2010bh associated with XRF 100316D (Cano et al. 2011b), SN~2006aj associated with XRF~060218 (Ferrero et al. 2006), as well as local SNe Ibc presented in C13.

Using the model in C13 we estimated the nickel mass, ejecta mass and kinetic energy of SN~2013fu.  Without knowledge of the peak photospheric velocity of SN~2013fu, we used the average peak photospheric velocity determined by C13 for a sample of GRB-SNe: $v_{ph}\approx20\pm2.5 \times 10^{3}$ km s$^{-1}$.  To estimate the nickel mass we computed the average luminosity factor from the $r$ and $i$ filters (neglecting the $z$ observation as it is not host subtracted and is therefore an overestimate of the SN's brightness), $\bar{k}=0.85\pm0.20$.  We estimated the ejecta velocity using the peak photospheric velocity and an average of the stretch factor in $R$ and $i$, $\bar{s}=0.83\pm0.02$.  We find bolometric properties of: $M_{\rm Ni}=0.30\pm0.07$ $M_{\sun}$, $M_{\rm ej}=4.71^{+0.79}_{-0.59}$ $M_{\odot}$ and $E_{\rm K}=1.87^{+0.90}_{-0.62} \times 10^{52}$ ergs.  The uncertainties in the ejecta mass and kinetic energy arise from the uncertainties in the stretch and luminosity factors as well as the spread of peak ejecta velocities around the mean value in C13.

\begin{table*}
\centering
\setlength{\tabcolsep}{2.0pt}
\setlength{\extrarowheight}{4.5pt}
\caption{GRB-SNe: Observational and Physical Properties ($UBVRIJH$ rest-frame wavelength range)}
\label{table:GRBSN_properties}
\begin{tabular}{cccccccccc}

\hline																			
GRB	&	SN	&	Filter (obs)	&	$k$	&	$s$	&	$T_{\rm peak,obs}$ (d)	&	$M_{\rm peak,obs}$	&	$M_{\rm Ni}$ ($\rm M_{\odot}$)	&	$M_{\rm ej}$ ($\rm M_{\odot}$)$^{\dagger}$	&	$\rm E_{K}$ ($10^{51}$ erg)$^{\dagger}$		\\
\hline
120729A	&	-	&	$r$	&	$1.29\pm0.19$	&	1.0 (fixed)	&	-	&	$-18.96\pm0.15$	&	-	&	-	&	-	\\
120729A	&	-	&	$i$	&	$0.76\pm0.11$	&	1.0 (fixed)	&	-	&	$-19.29\pm0.15$	&	-	&	-	&	-	\\
120729A	&	-	&	\textbf{average}	&	$1.02\pm0.26$	&	1.0 (fixed)	&	-	&	-	&	$0.42\pm0.11$	&	-	&	-	\\
\hline																			
130215A	&	2013ez	&	$r$	&	$0.6-0.75$	&	1.0 (fixed)	&	-	&	$-18.7$ to $-19.0$	&	-	&	-	&	-	\\
130215A	&	2013ez	&	$i$	&	$0.6-0.75$	&	1.0 (fixed)	&	-	&	$-19.0$ to $-19.3$	&	-	&	-	&	-	\\
130215A	&	2013ez	&	\textbf{average}	&	$0.6-0.75$	&	1.0 (fixed)	&	-	&	-	&	$0.25$ -- $0.30$	&	-	&	-	\\
\hline																			
130831A	&	2013fu	&	$r$	&	$0.65\pm0.03$	&	$0.84\pm0.03$	&	$18.60\pm0.67$	&	$-18.89\pm0.05$	&	-	&	-	&	-	\\
130831A	&	2013fu	&	$i$	&	$1.04\pm0.05$	&	$0.81\pm0.03$	&	$18.53\pm0.69$	&	$-19.56\pm0.05$	&	-	&	-	&	-	\\
130831A	&	2013fu	&	$z$	&	$1.02\pm0.19^{\ddagger}$	&	0.81 (fixed)	&	$\approx19.1$	&	$-19.47\pm0.19$	&	-	&	-	&	-	\\
130831A	&	2013fu	&	\textbf{average} ($r$ \& $i$)	&	$0.85\pm0.20$	&	$0.83\pm0.02$	&	-	&	-	&	$0.30\pm0.07$	&	$4.71^{+0.79}_{-0.59}$	&	$18.7^{+9.0}_{-6.2}$	\\
\hline																			
\hline																			
&		&	SN type	&	$M_{\rm Ni}$ ($\rm M_{\odot}$)	&	$M_{\rm Ej}$ ($\rm M_{\odot}$)	&	$\rm E_{K}$ ($10^{51}$ erg)	&		&		&		&		\\
\hline																			
&		&	Ib	&	0.16 ($\sigma=0.22$)	&	3.89 ($\sigma=2.77$)	&	2.3 ($\sigma=2.6$)	&		&		&		&		\\
&		&	Ic	&	0.19 ($\sigma=0.19$)	&	3.40 ($\sigma=4.51$)	&	2.2 ($\sigma=3.3$)	&		&	\textbf{median values from}	&		&		\\
&		&	Ibc	&	0.18 ($\sigma=0.21$)	&	3.56 ($\sigma=3.51$)	&	2.2 ($\sigma=2.8$)	&		&	\textbf{C13.}	&		&		\\
&		&	Ic-BL	&	0.26 ($\sigma=0.33$)	&	3.90 ($\sigma=3.44$)	&	10.9 ($\sigma=8.9$)	&		&		&		&		\\
&		&	GRB/XRF	&	0.34 ($\sigma=0.24$)	&	5.91 ($\sigma=3.87$)	&	22.3 ($\sigma=15.2$)	&		&		&		&		\\
\hline

\end{tabular}
\begin{flushleft}
$^{\dagger}$ Ejecta mass and kinetic energy are calculated using the average peak photospheric velocity of $v_{\rm ph}\approx(20\pm2.5) \times 10^{3}$ km s$^{-1}$ determined for a sample of seven GRB-SNe in C13.\\
$^{\ddagger}$ $z$ observations of GRB 130831A are not host-subtracted, and are not considered when calculating the average luminosity and stretch factors.
\end{flushleft}
\end{table*}

\section{Discussion \& Conclusions}
\label{sec:Discussion}

\begin{figure}
 \centering
 \includegraphics[bb=0 0 265 171, scale=0.93]{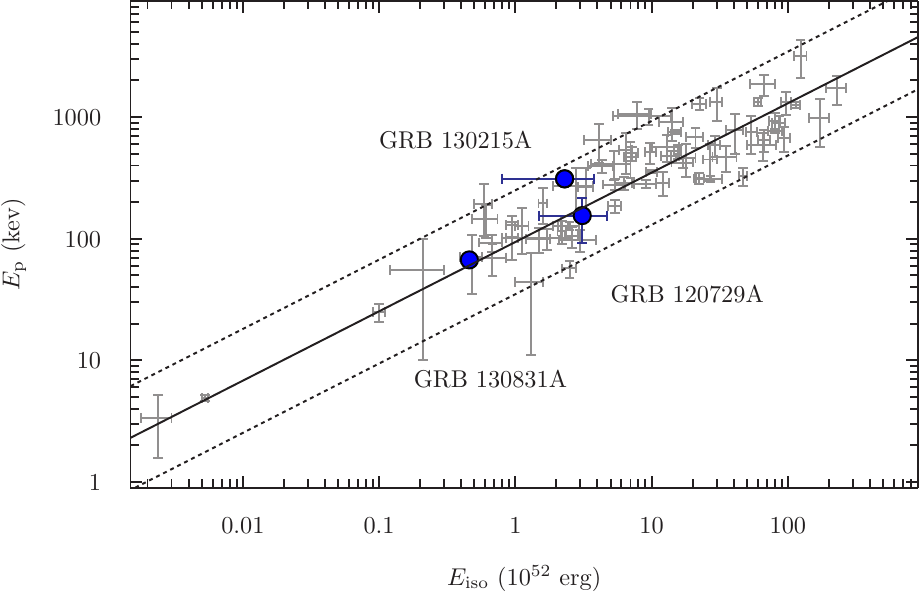}
 \caption{Our trio of GRBs on the $E_{\rm p}$ -- $E_{\rm iso, \gamma}$ plane.  The sample of GRBs presented in Amati et al. (2008) are shown in grey, along with their best fit to a single powerlaw ($\alpha=0.57$) and the 2-$\sigma$ uncertainty in their fit.  GRBs 120729A and 130831A lie perfectly on the Amati relation, while GRB 130215A is within the 2-$\sigma$ confidence interval.}
 \label{fig:amati}
\end{figure}

\subsection{The Supernovae}

We presented optical/NIR photometry for three GRB-SNe.  We derived the luminosity factor ($k$) for each SN, and the stretch factor ($s$) of SN~2013fu relative to a template supernova (SN~1998bw), which has been redshifted/k-corrected to that of each GRB-SN considered here.  We also estimated the peak, observer-frame magnitude of each SN in every available filter, as well as the time of peak light for SN~2013fu.  

We also presented a spectrum of SN~2013ez, which was associated with GRB~130215A.  Absorption features near 5100 \AA $ $ and 6200 \AA $ $ are seen, which are thought to be blueshifted \ion{Fe}{ii} $\lambda$5169 and \ion{Si}{ii} $\lambda$6355 respectively.  Using a Monte Carlo (MC) simulation we fit single and double Gaussians to these features, finding blueshifted velocities of $v_{Fe}\approx-4000\pm1500$ km s$^{-1}$ and $v_{Si}\approx-6350\pm800$ km s$^{-1}$.  These velocities agree within their respective errorbars.  

The spectrum was taken at rest-frame $t-t_{0}=16.1$d, which despite the unknown time of peak light, is likely to be no more than a few days from peak light if it evolved similarly to SN~1998bw, or $\sim$ a week from peak light if it evolved at a faster rate such as seen for SN~2006aj and SN~2010bh.  The blueshifted velocity seen in SN~2013ez is $\sim$8000--10,000 km s$^{-1}$ less than that of the slowest velocity of a sample of seven GRB/XRF-SNe investigated by Schulze et al. (2014) at the same rest-frame time.  The slowest blueshifted velocity of \ion{Fe}{ii} $\rm \lambda5169$ at $t-t_{0}=$16 d (rest-frame) in their sample is $\sim$14,000 km s$^{-1}$ for SN~2006aj.  Thus the narrow absorption features along with the slower blueshifted velocity of SN~2013ez makes it more appropriate to consider SN~2013ez as a type Ic rather than a type IcBL, making it the first GRB-SN to have this classification.  

The only other GRB-SN to have a possible classification of Ic was SN~2001lt, which was associated with GRB~021211 (Della Valle et al. 2003).  For this SN, blueshifted Ca II was seen at $\approx$14,400 km s$^{-1}$ at $t-t_{0}\approx$13.4 d (rest-frame), which, while being a lower value than seen in the majority of the other GRB-SNe, is comparable to SN~2006aj at a similar rest-frame epoch.  Moreover, the average peak photospheric velocity of a sample of seven SNe IcBL investigated by C13 is $\approx$15,000 km s$^{-1}$ ($\sigma=$4000 km s$^{-1}$), which is commensurate to that of SN~2001lt, making it comfortably classified as a type Ic-BL rather than Ic.

When analysing the optical properties of SN~2013fu, associated with GRB~130831A, we found that it is brighter in the redder filters: $k_{R}=0.65\pm0.03$, $k_{i}=1.04\pm0.05$ and $k_{z}=1.02\pm0.19$.\footnote{We note that a non-zero rest-frame extinction could partially explain this flux decrement.  In our second energy spectrum we find $A_{V,\rm rest}=0.21^{+0.28}_{-0.21}$ mag, which we argue implies zero rest-frame extinction.  However, we acknowledge that a non-zero extinction is still compatible with our result, which would partially alleviate this issue.}  If we take our result at face value, the red colour of 2013fu suggests that there is a suppression of flux in observer-frame \textit{R}-band ($\approx$ $B$-band in rest-frame, e.g. $\lambda_{\rm eff}=6588/(1+z)\approx4500$ \AA) due to metal line blanketing.  Line blanketing by \ion{Fe}{ii} and Ti II, which suppresses flux blueward of $\sim 4000$ \AA, was observed for Type Ib SN~1999dn (Branch et al. 2002; Deng et al. 2000; Cano et al. 2014).  Flux suppression due to several iron-group elements was also observed for SN~2006aj (Sollerman et al. 2006), while metal line blanketing was also suggested by Bloom et al. (1999) to explain the red colour of the SN bump of GRB~980326.

For each SN we used the luminosity factor averaged over all available filters to estimate the amount of nickel nucleosynthesized during the explosion, while for SN~2013fu we used the well-sampled LCs to estimate the ejecta mass and kinetic energy of the SN.  We do not have knowledge of the photospheric velocity of the SN at peak light, which is a necessary ingredient of the analytical model of Arnett (1982) to estimate $M_{ej}$ and $E_{K}$.  Instead we used the average photospheric velocity for a sample of GRB-SNe presented in C13.  A summary of the observational and physical properties of our three GRB-SNe are presented in Table \ref{table:GRBSN_properties}.

When estimating the bolometric properties of the GRB-SNe presented here, we have used the method described in C13, where a lengthly discussion of the caveats of this modelling can be found.  C13 determined the median bolometric properties of a sample of 20 GRB-SNe, finding: $M_{\rm Ni} \sim 0.3-0.35$ $\rm M_{\odot}$, $M_{\rm ej} \sim 6.0$~$\rm M_{\odot}$ and $E_{\rm K} \sim 2.0 \times 10^{52}$ erg.  The nickel masses derived here agree well with the range in C13, while the ejecta mass and kinetic energy of SN~2013fu is similar to those found in C13.  In terms of physical properties, the GRB-SNe in this paper are quite typical of other GRB-SNe.  In contrast, the nickel masses are much higher than those seen for SNe Ibc that are not associated with GRBs.  C13 derived the median nickel masses for the largest sample of SNe Ibc yet considered (32 Ibc and nine IcBL), finding: SNe Ibc: $M_{\rm Ni} \sim 0.15-0.18$ $\rm M_{\odot}$, and Ic-BL: $M_{\rm Ni} \sim 0.25$ ${\rm M_\odot}$.  Similarly the ejecta mass and kinetic energy of SN~2013fu is larger than those of the C13 sample of SNe Ibc ($M_{\rm ej} \sim 3.4-3.9$ $\rm M_{\odot}$, $E_{\rm K} \sim 0.2 \times 10^{52}$ erg) and Ic-BL ($M_{\rm ej} \sim 3.9$~$\rm M_{\odot}$, $E_{\rm K} \sim 1.0 \times 10^{52}$ erg).  

In Fig. \ref{fig:amati}, we have plotted these three GRBs in the $E_{\rm p}$ -- $E_{\rm iso,\gamma}$ plane.  We have used the sample of GRBs in Amati et al. (2008) and the values of $E_{\rm iso,\gamma}$ and $E_{\rm p}$ presented in the introduction of each GRB section.  Also plotted is a simple power-law, where we have used an index of $\alpha=0.57$, a normalization constant of $\Lambda=94$, and the 2-$\sigma$ confidence intervals in the Amati relation (Amati et al. 2008).  It is seen that GRBs 130831A and 120729A lie right on the fit, while GRB 130215A is within the upper bounds of the 2-$\sigma$ confidence interval.

Taking a step back, SN 2013ez is now the second known GRB-SN, after SN 2002lt, for which it is more likely that it is of type Ic rather than Ic-BL.  These results seem somewhat contrary to observations of all other GRB- and XRF-SNe, where unambiguous broad lines are seen in time-series spectra of the SNe.  Up to now, the general consensus has been that GRB/XRF-SNe are therefore a rare subclass of SNe Ic-BL, which themselves are a rare subclass of the general SNe Ibc population (e.g. Podsiadlowski et al. 2004; Guetta \& Della Valle 2007).  But are SN 2013ez and SN 2002lt really non-broad-lined SNe Ic? One way to reconcile this growing heterogeneity of GRB-SNe is to check whether the timing of observations affects how we spectroscopically classify each accompanying SN.  The spectrum of SN~2002lt was obtained at $t-t_{0}\approx$15 d (rest-frame), which is at a similar epoch as the timing of the GTC spectrum obtained of SN~2013ez.  The time of peak light of SN~2013ez is not well constrained, and could have been anywhere from 5--10 days after maximum light (rest-frame).  If the SN ejecta of these two events are less dense than expected, than the ejecta will become optically thin more rapidly than in the case of other GRB-SNe, and thus much more difficult to identify as a type Ic-BL.  An example of this is Ic-BL SN~2002ap (Mazzali et al. 2002) that showed broad features in the pre-maximum spectra which quickly vanished after maximum.  In their spectral modelling, the photospheric velocity rapidly decreased from $\sim$30,000 km s$^{-1}$ at two days after explosion, to $\sim$20,000 km s$^{-1}$ only three days later.  Therefore, it is not unfathomable that SN~2013ez may have had more rapid velocities earlier on, but they disappeared by the time we obtained our spectrum.  This clearly highlights the need to obtain more than one spectrum of the accompanying SN in order to more accurately quantify the line velocities and its spectral classification.

\subsection{The magnetar model}

In this work we derived the initial spin period and magnetic field of a possible millisecond magnetar central engine for GRB~130215A from optical/NIR observations.  We constructed a bolometric LC from contemporaneous $grizJH$ observations of GRB~130215A and fit it with the model from Zhang \& M\'esz\'aros (2001; see as well Rowlinson et al. 2013), finding $\rm P_{0}=12.0$ ms, a magnetic field strength of $\rm B=1.1 \times 10^{15}$ G, a plateau luminosity of $\rm L_{plat}=6.1 \times 10^{44}$ erg s$^{-1}$ and a plateau duration (rest-frame) of $\rm T_{plat}=2.3 \times 10^{5}$ s.  These values are not unrealistic, and reminiscent of those estimated for other long (GRB 060729: $P=$1.5--2.0 ms and $B=$0.25--3.2$ \times 10^{15}$ G, Xu et al. 2009; Dall'Osso et al. 2011; Lu \& Zhang 2014; and GRB~130427A: $P\sim20$ ms and $B\sim 10^{16}$ G, Bernardini et al. 2014), a sample of 10 long-duration GRBs in Lyons et al. (2010; their Table 2), and a sample of short GRBs detected by \emph{Swift} up to 2012 (see, e.g., Table 3 in Rowlinson et al. 2013).

We must consider the limitations of our data when interpreting this result.  The bolometric LC is constructed from observations of the AG+SN+host.  In early epochs the AG will dominate, however in the latter two epochs at $t-t_{0}=$8.1 and 9.8 d there will be some contribution of flux coming from SN~2013ez.   Moreover there will be a constant contribution from the underlying host galaxy in all epochs.  However, at early times the host and SN contribute a negligible amount of flux, though as the AG fades the SN becomes the dominant source of flux, which too fades, leaving the host as the only source of emission (this happens only after 100 days or more).  In our GTC images we have not detected the host to deep limits: $g>26.2$, $r>26.1$, $i>25.1$.  If the host had these magnitudes, it would contribute $\sim 3-8$, $25\%$ flux at $t-t_{0}=$ 9.8, 25 d respectively.  Obviously fainter magnitudes imply less host contribution.

\subsection{Future Prospects}

As discussed in the introduction, there are now almost a dozen spectroscopically associated GRB-SNe, though only a few have multi-band observations, with NIR observations severely lacking except in only the nearest GRB-SNe.  The bolometric properties of GRB-SNe have been shown to be statistically different to those of non-GRB SNe Ibc, implying that non-GRB SNe Ibc arise from different physical scenarios than GRB-SNe.  Without the possibility of directly detecting the progenitor star of a GRB-SNe, we must infer its properties indirectly via the application of advanced modelling techniques and simulations of the SN themselves.  Analytical models presented in e.g. Drout et al. (2011) and C13 can go only so far (within a factor of $\sim$2) in providing a clear description of the physical processes occurring during the SN, which themselves are highly dependent on the explosion mechanism, the evolutionary stage of the progenitor at the time of explosion, and the physical processes (and timescales of those processes) powering the SN.  In some cases the peak light may not be powered solely by nickel decay, but additionally by shock heating of the SN ejecta (e.g. Fryer et al. 2007).  This example highlights the need to employ more sophisticated modelling techniques that incorporate increasingly more of the physical processes that are occurring during the SN, as well as a more realistic description of the geometrical arrangement of the ejecta (symmetry vs. asymmetry).

As such there is still a great need for high quality optical and NIR photometry and spectra of GRB-SNe that are then modelled.  These can then be used to constrain the explosion mechanism and physical properties of the progenitor via SN modelling methods, such as Monte Carlo radiative transfer (RT) simulations (e.g. Mazzali \& Lucy 1993; Maeda et al. 2006; Kasen et al. 2006).  Simple RT simulations such as SYN++ (Thomas et al. 2011) provide a tool to approximately ascertain the chemical properties of the material passing through the photosphere at a given moment in time.  These results can then be used as input to a RT simulation.  The spectra can be used in a method such as ``abundance tomography'', which has been successfully used for SNe Ia (e.g. Hachinger et al. 2013; Mazzali et al. 2014) to determine the density structure and abundance stratification in the SN ejecta.  Massive stars are evolved and then exploded in hydrodynamic computer simulations, with the result being SN ejecta of a specific density structure and abundance stratification that can be directly compared with observations.  In this manner a focused observing strategy aimed at obtaining optical and NIR photometry and spectroscopy with 8--10-m class ground telescopes (to obtain rest-frame $BVRI$ LCs, and NIR if possible), which are combined with sophisticated simulations will undoubtedly provide deeper insight into the nature of the progenitor stars of GRB-SNe.

\section{Acknowledgements}

I am very grateful to Max de Pasquale for countless discussions regarding GRB physics, and Antonia Rowlinson for equally stimulating conversations regarding magnetars. We thank the referee for their insightful comments on the original manuscript.  

ZC gratefully acknowledges support by a Project Grant from the Icelandic Research Fund.
 
The Dark Cosmology Centre is funded by the Danish National Research Foundation.  

The research activity of AdUP, CT, RSR, and JGor is supported by Spanish research project AYA2012-39362-C02-02. J.Gor and RSR are also supported by project AYA2012-39727-C03-01.  AdUP acknowledges support by the European Commission under the Marie Curie Career Integration Grant programme (FP7-PEOPLE-2012-CIG 322307).

TK acknowledges support by the European Commission under the Marie Curie Intra-European Fellowship Programme.

The research activity of AM and PDA is supported by ASI grant INAF I/004/11/.

AP, AV acknowledge partial support by RFBR grants 12-02-01336, 14-02-10015, and AM, IM acknowledge support by RFBR grant 13-01-92204.

DAK acknowledges support by the Max-Planck Institut f\"ur Extraterrestrische Physik, Garching, and the Th\"uringer Landessternwarte Tautenburg.

JXP acknowledges support from NASA Swift Grant NNX13AJ67G.

JPUF acknowledges support from the ERC-StG grant EGGS-278202.

Based on observations made with the Nordic Optical Telescope, operated by the Nordic Optical Telescope Scientific Association at the Observatorio del Roque de los Muchachos, La Palma, Spain, of the Instituto de Astrofisica de Canarias.

Based on observations made with the Gran Telescopio Canarias (GTC), instaled in the Spanish Observatorio del Roque de los Muchachos of the Instituto de Astrofísica de Canarias, in the island of La Palma.

The Liverpool Telescope is operated by Liverpool John Moores University at the Observatorio del Roque de los Muchachos of the Instituto de Astrof\'{i}sica de Canarias. The Faulkes Telescopes are owned by Las Cumbres Observatory. CGM acknowledges support from the Royal Society, the Wolfson Foundation and the Science and Technology Facilities Council. 

The ROTSE-IIIb telescope is operated by Southern Methodist University at McDonald Observatory, Ft. Davis, Texas.

Parts of this research were conducted by the Australian Research Council Centre of Excellence for All-sky Astrophysics (CAASTRO), through project number CE110001020.

Additionally, we thank the RATIR project team and the staff of the Observatorio Astronómico Nacional on Sierra San Pedro M\'artir. RATIR is a collaboration between the University of California, the Universidad Nacional Auton\'oma de M\'exico, NASA Goddard Space Flight Center, and Arizona State University, benefiting from the loan of an H2RG detector and hardware and software support from Teledyne Scientific and Imaging. RATIR, the automation of the Harold L. Johnson Telescope of the Observatorio Astron\'omico Nacional on Sierra San Pedro M\'artir, and the operation of both are funded through NASA grants NNX09AH71G, NNX09AT02G, NNX10AI27G, and NNX12AE66G, CONACyT grants INFR-2009-01-122785 and CB-2008-101958, UNAM PAPIIT grant IN113810, and UC MEXUS-CONACyT grant CN 09-283.  

SS acknowledges support from CONICYT through FONDECYT grant 3140534, from Basal-CATA PFB-06/2007, Iniciativa Cientifica Milenio grant P10-064-F (Millennium Center for Supernova Science), and by Project IC120009 "Millennium Institute of Astrophysics (MAS)" of Iniciativa Cient\'ifica Milenio del Ministerio de Economía, Fomento y Turismo de Chile, with input from "Fondo de Innovaci\'{o}n para la Competitividad, del Ministerio de Econom\'{\i}a, Fomento y Turismo de Chile".

Part of this work is based on observations obtained at the Gemini Observatory, which is operated by the Association of Universities for Research in Astronomy, Inc., under a cooperative agreement with the NSF on behalf of the Gemini partnership: the National Science Foundation  (United States), the National Research Council (Canada), CONICYT (Chile), the Australian  Research Council (Australia), Minist\'{e}rio da Ci\^{e}ncia, Tecnologia e Inova\c{c}\~{a}o (Brazil) and Ministerio de Ciencia, Tecnolog\'{i}a e Innovaci\'{o}n Productiva (Argentina).

Part of the funding for GROND (both hardware as well as personnel) was generously granted from the Leibniz-Prize to Prof. G. Hasinger (DFG grant HA 1850/28-1).

\onecolumn
\centering
\setlength{\tabcolsep}{10.0pt}
\begin{longtable}{ccccccc} 
\caption{List of Photometry}\label{table:photometry}\\
\hline													
GRB	&	Filter	&	$t-t_{0}$ (d)	&	$m^{\dagger}$	&	$m_{err}$	&	System	&	Telescope	\\
\hline
\endhead
120729A	&	$g$	&	0.0053	&	16.60	&	0.09	&	AB	&	FTN	\\
120729A	&	$g$	&	0.0079	&	16.89	&	0.08	&	AB	&	FTN	\\
120729A	&	$g$	&	0.0114	&	17.27	&	0.18	&	AB	&	FTN	\\
120729A	&	$g$	&	0.0160	&	17.70	&	0.08	&	AB	&	FTN	\\
120729A	&	$g$	&	0.0226	&	18.02	&	0.10	&	AB	&	FTN	\\
120729A	&	$g$	&	0.0354	&	18.35	&	0.10	&	AB	&	FTN	\\
120729A	&	$g$	&	0.0564	&	19.16	&	0.08	&	AB	&	FTN	\\
120729A	&	$g$	&	0.0788	&	19.56	&	0.08	&	AB	&	FTN	\\
120729A	&	$g$	&	0.6564	&	25.14	&	0.67	&	AB	&	TNG	\\
120729A	&	$g$	&	0.7683	&	24.00	&	0.11	&	AB	&	GTC	\\
120729A	&	$g$	&	18.6095	&	24.44	&	0.06	&	AB	&	GTC	\\
120729A	&	$g$	&	24.6329	&	24.58	&	0.12	&	AB	&	GTC	\\
120729A	&	$g$	&	189.3991	&	24.44	&	0.15	&	AB	&	GTC	\\
120729A	&	$r$	&	0.7629	&	23.56	&	0.08	&	AB	&	GTC	\\
120729A	&	$r$	&	8.7604	&	24.11	&	0.11	&	AB	&	GTC	\\
120729A	&	$r$	&	18.5914	&	23.93	&	0.09	&	AB	&	GTC	\\
120729A	&	$r$	&	20.5592	&	23.97	&	0.09	&	AB	&	GTC	\\
120729A	&	$r$	&	189.3824	&	24.05	&	0.07	&	AB	&	GTC	\\
120729A	&	$i$	&	0.0068	&	15.47	&	0.06	&	AB	&	FTN	\\
120729A	&	$i$	&	0.0102	&	16.13	&	0.06	&	AB	&	FTN	\\
120729A	&	$i$	&	0.0142	&	16.44	&	0.06	&	AB	&	FTN	\\
120729A	&	$i$	&	0.0202	&	16.88	&	0.07	&	AB	&	FTN	\\
120729A	&	$i$	&	0.0282	&	17.06	&	0.06	&	AB	&	FTN	\\
120729A	&	$i$	&	0.0348	&	17.23	&	0.07	&	AB	&	FTN	\\
120729A	&	$i$	&	0.0427	&	17.55	&	0.07	&	AB	&	FTN	\\
120729A	&	$i$	&	0.0474	&	17.65	&	0.09	&	AB	&	FTN	\\
120729A	&	$i$	&	0.0515	&	17.87	&	0.08	&	AB	&	FTN	\\
120729A	&	$i$	&	0.0573	&	17.98	&	0.07	&	AB	&	FTN	\\
120729A	&	$i$	&	0.0654	&	18.24	&	0.07	&	AB	&	FTN	\\
120729A	&	$i$	&	0.0719	&	18.32	&	0.09	&	AB	&	FTN	\\
120729A	&	$i$	&	0.0799	&	18.52	&	0.08	&	AB	&	FTN	\\
120729A	&	$i$	&	0.6433	&	22.78	&	0.10	&	AB	&	TNG	\\
120729A	&	$i$	&	0.6447	&	22.52	&	0.35	&	AB	&	LT	\\
120729A	&	$i$	&	0.6468	&	22.64	&	0.10	&	AB	&	TNG	\\
120729A	&	$i$	&	0.7575	&	23.24	&	0.06	&	AB	&	GTC	\\
120729A	&	$i$	&	18.6275	&	23.66	&	0.07	&	AB	&	GTC	\\
120729A	&	$i$	&	24.6534	&	23.63	&	0.07	&	AB	&	GTC	\\
120729A	&	$i$	&	189.4159	&	23.89	&	0.07	&	AB	&	GTC	\\
120729A	&	$z$	&	0.7738	&	23.27	&	0.27	&	AB	&	GTC	\\
120729A	&	$z$	&	189.4331	&	23.86	&	0.14	&	AB	&	GTC	\\
120729A	&	$V$	&	0.0060	&	15.91	&	0.10	&	Vega	&	FTN	\\
120729A	&	$V$	&	0.5864	&	24.21	&	0.47	&	Vega	&	TNG	\\
120729A	&	$R$	&	0.0033	&	15.05	&	0.06	&	Vega	&	FTN	\\
120729A	&	$R$	&	0.0037	&	15.06	&	0.06	&	Vega	&	FTN	\\
120729A	&	$R$	&	0.0041	&	15.21	&	0.07	&	Vega	&	FTN	\\
120729A	&	$R$	&	0.0090	&	16.02	&	0.07	&	Vega	&	FTN	\\
120729A	&	$R$	&	0.0127	&	16.39	&	0.07	&	Vega	&	FTN	\\
120729A	&	$R$	&	0.0181	&	16.81	&	0.08	&	Vega	&	FTN	\\
120729A	&	$R$	&	0.0254	&	17.09	&	0.08	&	Vega	&	FTN	\\
120729A	&	$R$	&	0.0327	&	17.26	&	0.07	&	Vega	&	FTN	\\
120729A	&	$R$	&	0.0399	&	17.46	&	0.08	&	Vega	&	FTN	\\
120729A	&	$R$	&	0.0464	&	17.66	&	0.12	&	Vega	&	FTN	\\
120729A	&	$R$	&	0.0500	&	17.70	&	0.10	&	Vega	&	FTN	\\
120729A	&	$R$	&	0.0552	&	18.00	&	0.08	&	Vega	&	FTN	\\
120729A	&	$R$	&	0.0626	&	18.12	&	0.08	&	Vega	&	FTN	\\
120729A	&	$R$	&	0.0698	&	18.24	&	0.09	&	Vega	&	FTN	\\
120729A	&	$R$	&	0.0771	&	18.54	&	0.08	&	Vega	&	FTN	\\
120729A	&	$R$	&	0.0865	&	18.79	&	0.07	&	Vega	&	FTN	\\
120729A	&	$R$	&	0.5983	&	22.50	&	0.46	&	Vega	&	LT	\\
120729A	&	$R$	&	0.6301	&	23.23	&	0.18	&	Vega	&	TNG	\\
120729A	&	$I_{c}$	&	0.7069	&	$>$21.3	&	-	&	Vega	&	IAC80	\\
\hline
\hline
130215A	&	$g$	&	0.9783	&	20.84	&	0.06	&	AB	&	GROND	\\
130215A	&	$g$	&	2.9563	&	21.73	&	0.21	&	AB	&	GROND	\\
130215A	&	$g$	&	9.7970	&	22.76	&	0.23	&	AB	&	GTC	\\
130215A	&	$g$	&	372.8488	&	$>$26.2	&	-	&	AB	&	GTC	\\
130215A	&	$r$	&	0.00810	&	14.09	&	0.04	&	AB	&	ROTSE$^{\dagger}$	\\
130215A	&	$r$	&	0.00823	&	14.17	&	0.05	&	AB	&	ROTSE$^{\dagger}$	\\
130215A	&	$r$	&	0.00836	&	14.03	&	0.05	&	AB	&	ROTSE$^{\dagger}$	\\
130215A	&	$r$	&	0.00849	&	14.10	&	0.05	&	AB	&	ROTSE$^{\dagger}$	\\
130215A	&	$r$	&	0.00863	&	14.06	&	0.05	&	AB	&	ROTSE$^{\dagger}$	\\
130215A	&	$r$	&	0.00869	&	14.09	&	0.02	&	AB	&	ROTSE$^{\dagger}$	\\
130215A	&	$r$	&	0.00876	&	14.07	&	0.05	&	AB	&	ROTSE$^{\dagger}$	\\
130215A	&	$r$	&	0.00889	&	14.11	&	0.06	&	AB	&	ROTSE$^{\dagger}$	\\
130215A	&	$r$	&	0.00902	&	14.00	&	0.05	&	AB	&	ROTSE$^{\dagger}$	\\
130215A	&	$r$	&	0.00916	&	14.13	&	0.05	&	AB	&	ROTSE$^{\dagger}$	\\
130215A	&	$r$	&	0.00929	&	14.16	&	0.06	&	AB	&	ROTSE$^{\dagger}$	\\
130215A	&	$r$	&	0.00953	&	14.22	&	0.07	&	AB	&	ROTSE$^{\dagger}$	\\
130215A	&	$r$	&	0.00987	&	14.28	&	0.07	&	AB	&	ROTSE$^{\dagger}$	\\
130215A	&	$r$	&	0.0102	&	14.28	&	0.04	&	AB	&	ROTSE$^{\dagger}$	\\
130215A	&	$r$	&	0.0105	&	14.28	&	0.04	&	AB	&	ROTSE$^{\dagger}$	\\
130215A	&	$r$	&	0.0109	&	14.35	&	0.07	&	AB	&	ROTSE$^{\dagger}$	\\
130215A	&	$r$	&	0.0110	&	14.34	&	0.01	&	AB	&	ROTSE$^{\dagger}$	\\
130215A	&	$r$	&	0.0112	&	14.32	&	0.06	&	AB	&	ROTSE$^{\dagger}$	\\
130215A	&	$r$	&	0.0115	&	14.37	&	0.07	&	AB	&	ROTSE$^{\dagger}$	\\
130215A	&	$r$	&	0.0119	&	14.50	&	0.03	&	AB	&	ROTSE$^{\dagger}$	\\
130215A	&	$r$	&	0.0122	&	14.47	&	0.07	&	AB	&	ROTSE$^{\dagger}$	\\
130215A	&	$r$	&	0.0125	&	14.61	&	0.06	&	AB	&	ROTSE$^{\dagger}$	\\
130215A	&	$r$	&	0.0131	&	14.60	&	0.03	&	AB	&	ROTSE$^{\dagger}$	\\
130215A	&	$r$	&	0.0139	&	14.72	&	0.04	&	AB	&	ROTSE$^{\dagger}$	\\
130215A	&	$r$	&	0.0147	&	14.75	&	0.03	&	AB	&	ROTSE$^{\dagger}$	\\
130215A	&	$r$	&	0.0155	&	14.83	&	0.04	&	AB	&	ROTSE$^{\dagger}$	\\
130215A	&	$r$	&	0.0163	&	14.86	&	0.04	&	AB	&	ROTSE$^{\dagger}$	\\
130215A	&	$r$	&	0.0167	&	14.92	&	0.02	&	AB	&	ROTSE$^{\dagger}$	\\
130215A	&	$r$	&	0.0171	&	15.06	&	0.11	&	AB	&	ROTSE$^{\dagger}$	\\
130215A	&	$r$	&	0.0179	&	15.02	&	0.04	&	AB	&	ROTSE$^{\dagger}$	\\
130215A	&	$r$	&	0.0187	&	15.16	&	0.12	&	AB	&	ROTSE$^{\dagger}$	\\
130215A	&	$r$	&	0.0195	&	15.18	&	0.05	&	AB	&	ROTSE$^{\dagger}$	\\
130215A	&	$r$	&	0.0203	&	15.19	&	0.12	&	AB	&	ROTSE$^{\dagger}$	\\
130215A	&	$r$	&	0.0211	&	15.24	&	0.06	&	AB	&	ROTSE$^{\dagger}$	\\
130215A	&	$r$	&	0.0219	&	15.30	&	0.12	&	AB	&	ROTSE$^{\dagger}$	\\
130215A	&	$r$	&	0.0227	&	15.44	&	0.10	&	AB	&	ROTSE$^{\dagger}$	\\
130215A	&	$r$	&	0.0235	&	15.45	&	0.10	&	AB	&	ROTSE$^{\dagger}$	\\
130215A	&	$r$	&	0.0243	&	15.56	&	0.13	&	AB	&	ROTSE$^{\dagger}$	\\
130215A	&	$r$	&	0.0247	&	15.50	&	0.03	&	AB	&	ROTSE$^{\dagger}$	\\
130215A	&	$r$	&	0.0251	&	15.58	&	0.06	&	AB	&	ROTSE$^{\dagger}$	\\
130215A	&	$r$	&	0.0259	&	15.60	&	0.10	&	AB	&	ROTSE$^{\dagger}$	\\
130215A	&	$r$	&	0.0267	&	15.68	&	0.08	&	AB	&	ROTSE$^{\dagger}$	\\
130215A	&	$r$	&	0.0275	&	15.64	&	0.11	&	AB	&	ROTSE$^{\dagger}$	\\
130215A	&	$r$	&	0.0283	&	15.70	&	0.12	&	AB	&	ROTSE$^{\dagger}$	\\
130215A	&	$r$	&	0.0291	&	15.73	&	0.07	&	AB	&	ROTSE$^{\dagger}$	\\
130215A	&	$r$	&	0.0299	&	15.85	&	0.13	&	AB	&	ROTSE$^{\dagger}$	\\
130215A	&	$r$	&	0.0307	&	15.74	&	0.12	&	AB	&	ROTSE$^{\dagger}$	\\
130215A	&	$r$	&	0.0316	&	15.79	&	0.12	&	AB	&	ROTSE$^{\dagger}$	\\
130215A	&	$r$	&	0.0324	&	15.86	&	0.12	&	AB	&	ROTSE$^{\dagger}$	\\
130215A	&	$r$	&	0.0328	&	15.92	&	0.03	&	AB	&	ROTSE$^{\dagger}$	\\
130215A	&	$r$	&	0.0332	&	15.93	&	0.09	&	AB	&	ROTSE$^{\dagger}$	\\
130215A	&	$r$	&	0.0341	&	16.05	&	0.09	&	AB	&	ROTSE$^{\dagger}$	\\
130215A	&	$r$	&	0.0349	&	15.99	&	0.11	&	AB	&	ROTSE$^{\dagger}$	\\
130215A	&	$r$	&	0.0357	&	16.19	&	0.14	&	AB	&	ROTSE$^{\dagger}$	\\
130215A	&	$r$	&	0.0366	&	16.12	&	0.16	&	AB	&	ROTSE$^{\dagger}$	\\
130215A	&	$r$	&	0.0374	&	16.08	&	0.10	&	AB	&	ROTSE$^{\dagger}$	\\
130215A	&	$r$	&	0.0382	&	16.12	&	0.09	&	AB	&	ROTSE$^{\dagger}$	\\
130215A	&	$r$	&	0.0391	&	16.27	&	0.10	&	AB	&	ROTSE$^{\dagger}$	\\
130215A	&	$r$	&	0.0399	&	16.10	&	0.12	&	AB	&	ROTSE$^{\dagger}$	\\
130215A	&	$r$	&	0.0407	&	16.22	&	0.17	&	AB	&	ROTSE$^{\dagger}$	\\
130215A	&	$r$	&	0.0412	&	16.21	&	0.04	&	AB	&	ROTSE$^{\dagger}$	\\
130215A	&	$r$	&	0.0416	&	16.20	&	0.09	&	AB	&	ROTSE$^{\dagger}$	\\
130215A	&	$r$	&	0.0424	&	16.23	&	0.16	&	AB	&	ROTSE$^{\dagger}$	\\
130215A	&	$r$	&	0.0432	&	16.21	&	0.11	&	AB	&	ROTSE$^{\dagger}$	\\
130215A	&	$r$	&	0.0441	&	16.19	&	0.11	&	AB	&	ROTSE$^{\dagger}$	\\
130215A	&	$r$	&	0.0449	&	16.18	&	0.10	&	AB	&	ROTSE$^{\dagger}$	\\
130215A	&	$r$	&	0.0457	&	16.30	&	0.15	&	AB	&	ROTSE$^{\dagger}$	\\
130215A	&	$r$	&	0.0466	&	16.27	&	0.11	&	AB	&	ROTSE$^{\dagger}$	\\
130215A	&	$r$	&	0.0474	&	16.50	&	0.22	&	AB	&	ROTSE$^{\dagger}$	\\
130215A	&	$r$	&	0.0482	&	16.53	&	0.15	&	AB	&	ROTSE$^{\dagger}$	\\
130215A	&	$r$	&	0.0491	&	16.45	&	0.18	&	AB	&	ROTSE$^{\dagger}$	\\
130215A	&	$r$	&	0.0495	&	16.44	&	0.04	&	AB	&	ROTSE$^{\dagger}$	\\
130215A	&	$r$	&	0.0499	&	16.28	&	0.13	&	AB	&	ROTSE$^{\dagger}$	\\
130215A	&	$r$	&	0.0507	&	16.37	&	0.13	&	AB	&	ROTSE$^{\dagger}$	\\
130215A	&	$r$	&	0.0516	&	16.64	&	0.18	&	AB	&	ROTSE$^{\dagger}$	\\
130215A	&	$r$	&	0.0658	&	16.81	&	0.06	&	AB	&	ROTSE$^{\dagger}$	\\
130215A	&	$r$	&	0.0674	&	16.99	&	0.06	&	AB	&	RATIR	\\
130215A	&	$r$	&	0.0710	&	16.84	&	0.14	&	AB	&	ROTSE$^{\dagger}$	\\
130215A	&	$r$	&	0.0754	&	17.15	&	0.04	&	AB	&	RATIR	\\
130215A	&	$r$	&	0.0765	&	17.15	&	0.08	&	AB	&	RATIR	\\
130215A	&	$r$	&	0.0824	&	17.19	&	0.06	&	AB	&	RATIR	\\
130215A	&	$r$	&	0.0833	&	17.22	&	0.06	&	AB	&	RATIR	\\
130215A	&	$r$	&	0.0893	&	17.33	&	0.08	&	AB	&	RATIR	\\
130215A	&	$r$	&	0.0928	&	17.41	&	0.09	&	AB	&	RATIR	\\
130215A	&	$r$	&	0.0938	&	17.48	&	0.04	&	AB	&	RATIR	\\
130215A	&	$r$	&	0.0952	&	17.42	&	0.03	&	AB	&	RATIR	\\
130215A	&	$r$	&	0.0963	&	17.31	&	0.07	&	AB	&	RATIR	\\
130215A	&	$r$	&	0.0973	&	17.43	&	0.06	&	AB	&	RATIR	\\
130215A	&	$r$	&	0.1021	&	17.49	&	0.04	&	AB	&	RATIR	\\
130215A	&	$r$	&	0.1056	&	17.52	&	0.03	&	AB	&	RATIR	\\
130215A	&	$r$	&	0.1066	&	17.51	&	0.04	&	AB	&	RATIR	\\
130215A	&	$r$	&	0.1102	&	17.68	&	0.05	&	AB	&	RATIR	\\
130215A	&	$r$	&	0.1111	&	17.84	&	0.06	&	AB	&	RATIR	\\
130215A	&	$r$	&	0.1129	&	17.58	&	0.02	&	AB	&	RATIR	\\
130215A	&	$r$	&	0.1146	&	17.55	&	0.04	&	AB	&	RATIR	\\
130215A	&	$r$	&	0.1202	&	17.85	&	0.09	&	AB	&	RATIR	\\
130215A	&	$r$	&	0.1362	&	17.70	&	0.22	&	AB	&	ROTSE$^{\dagger}$	\\
130215A	&	$r$	&	0.1426	&	17.57	&	0.19	&	AB	&	ROTSE$^{\dagger}$	\\
130215A	&	$r$	&	0.1506	&	17.71	&	0.20	&	AB	&	ROTSE$^{\dagger}$	\\
130215A	&	$r$	&	0.1588	&	17.72	&	0.20	&	AB	&	ROTSE$^{\dagger}$	\\
130215A	&	$r$	&	0.1667	&	17.55	&	0.23	&	AB	&	ROTSE$^{\dagger}$	\\
130215A	&	$r$	&	0.9783	&	20.42	&	0.04	&	AB	&	GROND	\\
130215A	&	$r$	&	2.0713	&	21.08	&	0.13	&	AB	&	RATIR	\\
130215A	&	$r$	&	2.9563	&	21.19	&	0.09	&	AB	&	GROND	\\
130215A	&	$r$	&	3.1015	&	21.21	&	0.07	&	AB	&	RATIR	\\
130215A	&	$r$	&	4.0876	&	21.51	&	0.10	&	AB	&	RATIR	\\
130215A	&	$r$	&	5.7866	&	21.54	&	0.09	&	AB	&	NOT	\\
130215A	&	$r$	&	8.0828	&	22.59	&	0.18	&	AB	&	RATIR	\\
130215A	&	$r$	&	9.7887	&	22.40	&	0.10	&	AB	&	GTC	\\
130215A	&	$r$	&	13.8012	&	23.38	&	0.07	&	AB	&	NOT	\\
130215A	&	$r$	&	17.5687	&	23.81	&	0.42	&	AB	&	RATIR	\\
130215A	&	$r$	&	372.8629	&	$>$26.1	&	-	&	AB	&	GTC	\\
130215A	&	$i$	&	0.0674	&	16.67	&	0.06	&	AB	&	RATIR	\\
130215A	&	$i$	&	0.0685	&	16.61	&	0.17	&	AB	&	RATIR	\\
130215A	&	$i$	&	0.0730	&	16.79	&	0.10	&	AB	&	RATIR	\\
130215A	&	$i$	&	0.0754	&	16.89	&	0.04	&	AB	&	RATIR	\\
130215A	&	$i$	&	0.0765	&	16.83	&	0.08	&	AB	&	RATIR	\\
130215A	&	$i$	&	0.0824	&	17.00	&	0.07	&	AB	&	RATIR	\\
130215A	&	$i$	&	0.0833	&	17.06	&	0.07	&	AB	&	RATIR	\\
130215A	&	$i$	&	0.0883	&	17.06	&	0.07	&	AB	&	RATIR	\\
130215A	&	$i$	&	0.0893	&	17.04	&	0.08	&	AB	&	RATIR	\\
130215A	&	$i$	&	0.0928	&	17.08	&	0.07	&	AB	&	RATIR	\\
130215A	&	$i$	&	0.0938	&	17.22	&	0.03	&	AB	&	RATIR	\\
130215A	&	$i$	&	0.0952	&	17.17	&	0.03	&	AB	&	RATIR	\\
130215A	&	$i$	&	0.0963	&	17.16	&	0.06	&	AB	&	RATIR	\\
130215A	&	$i$	&	0.0973	&	17.25	&	0.07	&	AB	&	RATIR	\\
130215A	&	$i$	&	0.1021	&	17.26	&	0.04	&	AB	&	RATIR	\\
130215A	&	$i$	&	0.1056	&	17.26	&	0.03	&	AB	&	RATIR	\\
130215A	&	$i$	&	0.1066	&	17.25	&	0.04	&	AB	&	RATIR	\\
130215A	&	$i$	&	0.1102	&	17.34	&	0.05	&	AB	&	RATIR	\\
130215A	&	$i$	&	0.1111	&	17.29	&	0.05	&	AB	&	RATIR	\\
130215A	&	$i$	&	0.1129	&	17.31	&	0.02	&	AB	&	RATIR	\\
130215A	&	$i$	&	0.1146	&	17.38	&	0.05	&	AB	&	RATIR	\\
130215A	&	$i$	&	0.1202	&	17.51	&	0.09	&	AB	&	RATIR	\\
130215A	&	$i$	&	0.8175	&	20.04	&	0.05	&	AB	&	NOT	\\
130215A	&	$i$	&	0.9783	&	20.14	&	0.06	&	AB	&	GROND	\\
130215A	&	$i$	&	2.0713	&	20.98	&	0.14	&	AB	&	RATIR	\\
130215A	&	$i$	&	2.9563	&	20.82	&	0.12	&	AB	&	GROND	\\
130215A	&	$i$	&	3.1039	&	20.88	&	0.07	&	AB	&	RATIR	\\
130215A	&	$i$	&	4.0543	&	20.88	&	0.08	&	AB	&	RATIR	\\
130215A	&	$i$	&	4.0876	&	20.81	&	0.06	&	AB	&	RATIR	\\
130215A	&	$i$	&	4.1206	&	20.68	&	0.07	&	AB	&	RATIR	\\
130215A	&	$i$	&	4.7809	&	20.86	&	0.05	&	AB	&	NOT	\\
130215A	&	$i$	&	5.7932	&	21.07	&	0.11	&	AB	&	NOT	\\
130215A	&	$i$	&	8.0821	&	22.25	&	0.16	&	AB	&	RATIR	\\
130215A	&	$i$	&	9.7910	&	22.39	&	0.12	&	AB	&	GTC	\\
130215A	&	$i$	&	11.0831	&	22.68	&	0.26	&	AB	&	RATIR	\\
130215A	&	$i$	&	13.7889	&	22.78	&	0.08	&	AB	&	NOT	\\
130215A	&	$i$	&	17.5687	&	23.52	&	0.40	&	AB	&	RATIR	\\
130215A	&	$i$	&	24.2781	&	23.63	&	0.30	&	AB	&	NOT	\\
130215A	&	$i$	&	372.8358	&	$>$25.1	&	-	&	AB	&	GTC	\\
130215A	&	$z$	&	0.0671	&	16.35	&	0.20	&	AB	&	RATIR	\\
130215A	&	$z$	&	0.0785	&	16.58	&	0.08	&	AB	&	RATIR	\\
130215A	&	$z$	&	0.0842	&	16.64	&	0.07	&	AB	&	RATIR	\\
130215A	&	$z$	&	0.0890	&	16.71	&	0.07	&	AB	&	RATIR	\\
130215A	&	$z$	&	0.0904	&	17.09	&	0.13	&	AB	&	RATIR	\\
130215A	&	$z$	&	0.0938	&	17.07	&	0.10	&	AB	&	RATIR	\\
130215A	&	$z$	&	0.0959	&	16.98	&	0.04	&	AB	&	RATIR	\\
130215A	&	$z$	&	0.0972	&	17.01	&	0.06	&	AB	&	RATIR	\\
130215A	&	$z$	&	0.0980	&	17.15	&	0.10	&	AB	&	RATIR	\\
130215A	&	$z$	&	0.1029	&	17.02	&	0.05	&	AB	&	RATIR	\\
130215A	&	$z$	&	0.1077	&	17.07	&	0.06	&	AB	&	RATIR	\\
130215A	&	$z$	&	0.1119	&	17.10	&	0.07	&	AB	&	RATIR	\\
130215A	&	$z$	&	0.1139	&	17.11	&	0.04	&	AB	&	RATIR	\\
130215A	&	$z$	&	0.1212	&	17.18	&	0.11	&	AB	&	RATIR	\\
130215A	&	$z$	&	0.8296	&	20.03	&	0.04	&	AB	&	NOT	\\
130215A	&	$z$	&	0.9783	&	19.92	&	0.07	&	AB	&	GROND	\\
130215A	&	$z$	&	2.0719	&	20.46	&	0.27	&	AB	&	RATIR	\\
130215A	&	$z$	&	2.9563	&	20.74	&	0.12	&	AB	&	GROND	\\
130215A	&	$z$	&	3.1050	&	20.59	&	0.09	&	AB	&	RATIR	\\
130215A	&	$z$	&	4.0855	&	20.63	&	0.09	&	AB	&	RATIR	\\
130215A	&	$z$	&	5.8026	&	20.72	&	0.09	&	AB	&	NOT	\\
130215A	&	$z$	&	8.0838	&	21.62	&	0.22	&	AB	&	RATIR	\\
130215A	&	$z$	&	9.7940	&	22.27	&	0.14	&	AB	&	GTC	\\
130215A	&	$Y$	&	2.0719	&	20.30	&	0.25	&	AB	&	RATIR	\\
130215A	&	$Y$	&	3.1050	&	20.18	&	0.09	&	AB	&	RATIR	\\
130215A	&	$Y$	&	4.0855	&	20.33	&	0.12	&	AB	&	RATIR	\\
130215A	&	$Y$	&	8.0838	&	21.14	&	0.27	&	AB	&	RATIR	\\
130215A	&	$J$	&	0.0754	&	16.38	&	0.10	&	AB	&	RATIR	\\
130215A	&	$J$	&	0.0842	&	16.29	&	0.10	&	AB	&	RATIR	\\
130215A	&	$J$	&	0.0890	&	16.41	&	0.09	&	AB	&	RATIR	\\
130215A	&	$J$	&	0.0904	&	16.71	&	0.15	&	AB	&	RATIR	\\
130215A	&	$J$	&	0.0939	&	16.31	&	0.10	&	AB	&	RATIR	\\
130215A	&	$J$	&	0.0941	&	16.65	&	0.04	&	AB	&	RATIR	\\
130215A	&	$J$	&	0.0959	&	16.56	&	0.04	&	AB	&	RATIR	\\
130215A	&	$J$	&	0.0980	&	16.64	&	0.10	&	AB	&	RATIR	\\
130215A	&	$J$	&	0.1028	&	16.66	&	0.06	&	AB	&	RATIR	\\
130215A	&	$J$	&	0.1077	&	16.97	&	0.08	&	AB	&	RATIR	\\
130215A	&	$J$	&	0.1119	&	16.71	&	0.08	&	AB	&	RATIR	\\
130215A	&	$J$	&	0.1139	&	16.79	&	0.07	&	AB	&	RATIR	\\
130215A	&	$J$	&	0.1212	&	16.82	&	0.12	&	AB	&	RATIR	\\
130215A	&	$J$	&	0.9786	&	19.39	&	0.15	&	AB	&	GROND	\\
130215A	&	$J$	&	2.0720	&	20.47	&	0.41	&	AB	&	RATIR	\\
130215A	&	$J$	&	2.9611	&	20.08	&	0.26	&	AB	&	GROND	\\
130215A	&	$J$	&	3.1050	&	20.06	&	0.09	&	AB	&	RATIR	\\
130215A	&	$J$	&	4.0883	&	20.14	&	0.14	&	AB	&	RATIR	\\
130215A	&	$J$	&	8.0842	&	21.55	&	0.50	&	AB	&	RATIR	\\
130215A	&	$J$	&	331.8415	&	$>$23.2	&	-	&	Vega	&	CAHA	\\
130215A	&	$H$	&	0.0941	&	15.48	&	0.17	&	AB	&	RATIR	\\
130215A	&	$H$	&	0.9786	&	18.97	&	0.13	&	AB	&	GROND	\\
130215A	&	$H$	&	2.0720	&	19.54	&	0.19	&	AB	&	RATIR	\\
130215A	&	$H$	&	2.9611	&	19.94	&	0.31	&	AB	&	GROND	\\
130215A	&	$H$	&	3.1050	&	19.79	&	0.10	&	AB	&	RATIR	\\
130215A	&	$H$	&	4.0883	&	19.78	&	0.13	&	AB	&	RATIR	\\
130215A	&	$K$	&	0.9786	&	18.63	&	0.22	&	AB	&	GROND	\\
130215A	&	$K$	&	2.9611	&	18.93	&	0.23	&	AB	&	GROND	\\
\hline
\hline
130831A	&	$B$	&	2.0525	&	22.700	&	0.070	&	Vega	&	MAO	\\
130831A	&	$R_{c}$	&	2.0411	&	22.200	&	0.090	&	Vega	&	MAO	\\
130831A	&	$R_{c}$	&	3.3168	&	22.200	&	0.110	&	Vega	&	Mondy	\\
130831A	&	$R_{c}$	&	4.1061	&	22.530	&	0.090	&	Vega	&	Mondy	\\
130831A	&	$R_{c}$	&	4.2555	&	22.530	&	0.070	&	Vega	&	MAO	\\
130831A	&	$R_{c}$	&	5.2440	&	22.880	&	0.130	&	Vega	&	CrAO	\\
130831A	&	$R_{c}$	&	5.2658	&	22.870	&	0.060	&	Vega	&	MAO	\\
130831A	&	$R_{c}$	&	7.0446	&	22.990	&	0.080	&	Vega	&	MAO	\\
130831A	&	$R_{c}$	&	8.0586	&	22.900	&	0.060	&	Vega	&	MAO	\\
130831A	&	$R_{c}$	&	9.2558	&	23.110	&	0.060	&	Vega	&	MAO	\\
130831A	&	$R_{c}$	&	10.2526	&	22.940	&	0.080	&	Vega	&	MAO	\\
130831A	&	$R_{c}$	&	11.2419	&	22.930	&	0.110	&	Vega	&	MAO	\\
130831A	&	$R_{c}$	&	15.2202	&	22.870	&	0.080	&	Vega	&	MAO	\\
130831A	&	$R_{c}$	&	16.2715	&	22.770	&	0.070	&	Vega	&	MAO	\\
130831A	&	$R_{c}$	&	22.0721	&	22.530	&	0.290	&	Vega	&	MAO	\\
130831A	&	$R_{c}$	&	27.0558	&	22.900	&	0.170	&	Vega	&	MAO	\\
130831A	&	$R_{c}$	&	28.6928	&	22.910	&	0.094	&	Vega	&	WHT	\\
130831A	&	$R_{c}$	&	30.15945	&	23.360	&	0.160	&	Vega	&	MAO	\\
130831A	&	$R_{c}$	&	37.1378	&	23.590	&	0.160	&	Vega	&	MAO	\\
130831A	&	$R_{c}$	&	40.60034	&	23.640	&	0.230	&	Vega	&	Mondy	\\
130831A	&	$R_{c}$	&	44.5578	&	$>23.2$	&	-	&	Vega	&	Mondy	\\
130831A	&	$R_{c}$	&	66.3495	&	23.650	&	0.150	&	Vega	&	CrAO	\\
130831A	&	$R_{c}$	&	96.0754	&	23.750	&	0.350	&	Vega	&	Mondy	\\
130831A	&	$I_{c}$	&	28.6701	&	22.076	&	0.129	&	Vega	&	WHT	\\
130831A	&	$g$	&	0.3834	&	19.672	&	0.022	&	AB	&	NOT	\\
130831A	&	$g$	&	4.6225	&	23.464	&	0.034	&	AB	&	WHT	\\
130831A	&	$r$	&	0.3907	&	19.324	&	0.026	&	AB	&	NOT	\\
130831A	&	$r$	&	0.6974	&	20.423	&	0.025	&	AB	&	NOT	\\
130831A	&	$r$	&	0.7013	&	20.432	&	0.027	&	AB	&	NOT	\\
130831A	&	$r$	&	7.5947	&	22.995	&	0.048	&	AB	&	NOT	\\
130831A	&	$r$	&	13.4525	&	23.038	&	0.094	&	AB	&	NOT	\\
130831A	&	$r$	&	27.3877	&	23.414	&	0.055	&	AB	&	NOT	\\
130831A	&	$r$	&	68.5157	&	23.922	&	0.051	&	AB	&	Gemini	\\
130831A	&	$r$	&	156.3151	&	24.063	&	0.089	&	AB	&	NOT	\\
130831A	&	$i$	&	0.3901	&	19.100	&	0.027	&	AB	&	NOT	\\
130831A	&	$i$	&	1.6584	&	21.741	&	0.066	&	AB	&	NOT	\\
130831A	&	$i$	&	3.6759	&	22.403	&	0.053	&	AB	&	NOT	\\
130831A	&	$i$	&	4.6024	&	22.540	&	0.044	&	AB	&	WHT	\\
130831A	&	$i$	&	13.4761	&	22.580	&	0.061	&	AB	&	NOT	\\
130831A	&	$i$	&	22.6051	&	22.422	&	0.058	&	AB	&	NOT	\\
130831A	&	$i$	&	27.4342	&	22.843	&	0.054	&	AB	&	NOT	\\
130831A	&	$i$	&	68.5281	&	23.777	&	0.089	&	AB	&	Gemini	\\
130831A	&	$i$	&	127.2841	&	24.235	&	0.097	&	AB	&	LT	\\
130831A	&	$z$	&	0.3987	&	19.026	&	0.043	&	AB	&	NOT	\\
130831A	&	$z$	&	1.4975	&	21.218	&	0.099	&	AB	&	LT	\\
130831A	&	$z$	&	6.5690	&	22.720	&	0.112	&	AB	&	LT	\\
130831A	&	$z$	&	17.5640	&	22.726	&	0.200	&	AB	&	LT	\\

\hline

\caption{$^{\dagger}$ Apparent magnitudes of the AG+SN+host are not corrected for foreground or rest-frame extinction.\\
$^{\ddagger}$ROTSE magnitudes are unfiltered calibrated to $r$.}
\end{longtable}

\label{lastpage}

\end{document}